\documentclass{amsart}
\usepackage[foot]{amsaddr}
\usepackage{epsfig}
\usepackage{framed}
\usepackage{geometry}
\usepackage{soul}
\usepackage{amssymb,xcolor}
\usepackage{amsbsy,todonotes,cancel, mathrsfs, amsmath,mathtools}
\usepackage{amsthm}
\usepackage[colorlinks=true, pdfstartview=FitV, linkcolor=blue, citecolor=blue, urlcolor=blue]{hyperref}
\usetikzlibrary{calc}
\usetikzlibrary{decorations.markings}
\usepackage{verbatim}

\usepackage{ulem}    

\usepackage{booktabs, array} 
\newcolumntype{L}{>{$}l<{$}} 
\newcolumntype{C}{>{$}c<{$}} 

\renewcommand{\theequation}{\arabic{section}.\arabic{equation}}
\makeatletter
\@addtoreset{equation}{section}
\makeatother
\def\s{\sigma}

\def \G{\Gamma}

\def \scr{\mathscr}
\theoremstyle{plain}
\newtheorem{theorem}{Theorem}[section]
\newtheorem{question}[theorem]{Question}
\newtheorem{proposition}[theorem]{Proposition}
\newtheorem{corollary}[theorem]{Corollary}
\newtheorem{lemma}[theorem]{Lemma}

\theoremstyle{definition}
\newtheorem{definition}[theorem]{Definition}
\newtheorem{example}[theorem]{Example}

\newtheorem{exercise}[theorem]{Exercise}

\newtheorem{examps}[theorem]{Examples}

\newtheorem{remark}[theorem]{Remark}

\def \nn{\nonumber}
\definecolor {darkgreen}{rgb}{0,0.6,0}
\def\le{\left}
\def\eqref#1{(\ref{#1})}
\def\ri{\right}

\def \QED{\hfill $\blacksquare$\par \vskip 5pt}
\def\ds{\displaystyle}
\def\un{\underline}
\def\res{\mathop{\mathrm {res}}\limits_}

\definecolor{shadecolor}{rgb}{0.95, 0.95, 0.86}

\def\br{\begin{remark}}
\def\er{\end{remark}}
\def\bt{
\definecolor{shadecolor}{rgb}{0.95, 0.95, 0.86}
\begin{shaded}
\begin{theorem}}
\def\et{\end{theorem}
\end{shaded}}
\def\bq{
\definecolor{shadecolor}{rgb}{0.95, 0.95, 0.46}
\begin{shaded}
\begin{question}}
\def\eq{\end{question}
\end{shaded}}
\def\bd{
\begin{definition}}
\def\ed{\end{definition}
}
\def \Re{\operatorname{Re}}
\def \Im{\operatorname{Im}}
\def\bp{
\begin{shaded}
\begin{proposition}}
\def\ep{\end{proposition}
\end{shaded}}

\def\bc{\begin{corollary}}
\def\ec{\end{corollary}}

\def\brs{\begin{remarks}
\begin{enumerate}}

\def\ers{\end{enumerate}\end{remarks}}
\def\bx{\begin{example}\small}
\def\ex{\end{example}}
\def\bxr{\begin{exercise}\small}
\def\exr{\end{exercise}}
\def\bl{\begin{lemma}}
\def\el{\end{lemma}}
\def\bxs{\begin{examps}. \rm\begin{enumerate}}
\def\exs{\end{enumerate}\end{examps}}

\def\&{\hspace{-15pt}&}
\def\bea#1\eea{\begin{align}#1\end{align}}
\def\beas{\begin{eqnarray*}}
\def\eeas{\end{eqnarray*}}
\def \pa{\partial}
\def\C{{\mathbb C}}

\def\R{{\mathbb R}}

\def\N{{\mathbb N}}

\def\Z{{\mathbb Z}}

\def\d{\mathrm d}

\def\l{\lambda}

\def\A{\mathfrak A}

\def\1{{\bf 1}}

\def\wt{\widetilde}
\def\ds{\displaystyle}

\def \be#1\ee{\begin{align}#1\end{align}}

\def \bs{\boldsymbol}

\def\l{\lambda}
\def\o{\omega}
\def\Rscr{\mathcal R}

\def\eps{\varepsilon}
\def\hf{\frac 12}
\def\trho{\tilde\rho}

\newcommand{\transpose}{\intercal}

\begin{document}

\title[Partial degeneration of finite gap solutions of KdV]{Partial degeneration of  finite gap solutions to the Korteweg-de Vries equation: soliton  gas and scattering on elliptic backgrounds}

\author{M. Bertola$^{\spadesuit, \clubsuit, \heartsuit, \diamondsuit}$ }
\address{$^\spadesuit$ SISSA, International School for Advanced Studies, via Bonomea 265, Trieste, Italy }
\address{$^\clubsuit$Istituto Nazionale di Fisica Nucleare (INFN)}
\address{$^\heartsuit$Centre de recherches math\'ematiques,
Universit\'e de Montr\'eal\\ C.~P.~6128, succ. centre ville, Montr\'eal,
Qu\'ebec, Canada H3C 3J7}
\address{$^\diamondsuit$Department of Mathematics and Statistics, Concordia University, \\ 1455 de Maisonnueve W.,  Montr\'eal,
Qu\'ebec, Canada H3G 1M8}
\email{marco.bertola@concordia.ca}
\author{R. Jenkins$^\star$ and A. Tovbis$^\star$}
\address{$^{\star}$Department of Mathematics, University of Central Florida, Orlando, FL, USA}
\email{robert.jenkins@ucf.edu}
\email{alexander.tovbis@ucf.edu}

\email

\begin{abstract}
We obtain Fredholm type formulas for partial  degenerations of  Theta functions on (irreducible)
nodal curves of arbitrary genus, with emphasis on nodal curves of genus one. 
An application is the study of ``many-soliton" solutions on an elliptic (cnoidal) background standing wave for the Korteweg-de Vries (KdV) equation starting from a formula that is reminiscent of the classical  Kay-Moses formula for $N$-solitons. In particular, we represent such a solution as  a sum of the following two terms: a ``shifted" elliptic (cnoidal) background wave and a Kay-Moses type determinant
containing Jacobi theta functions for the solitonic content, which can be viewed as a collection of  solitary disturbances 
on the cnoidal background.
The expressions for the travelling (group) speed of these solitary disturbances, 
as well as   
for the interaction kernel describing the scattering of pairs of such solitary disturbances,
 are obtained explicitly in terms of Jacobi theta functions. 
We also show that genus $N+1$ finite gap solutions with random initial phases converge in probability to the deterministic cnoidal wave solution as   $N$ {bands} degenerate to a nodal curve  of genus one. 
 Finally,
we derive {the nonlinear dispersion relations and 
the equation of states  for
 the KdV soliton gas on  the} residual  elliptic background.
\end{abstract}


\maketitle

\tableofcontents

\section{Introduction and results}
The Korteweg-de Vries (KdV) equation
\be\label{KdV-i}
u_t + u_{xxx} + 6 u u_x=0, \ \ u = u(x,t)
\ee
is historically the first equation shown \cite{ZabKrus} to admit solitary waves. The simplest such solution is the single ``soliton" (solitary wave) solution 
\be
\label{singlesolitongenus0}
u(x,t) = \frac{|b|/2}{\cosh^2 \left( \frac{\sqrt{|b|}x - |b|^\frac 32 t + \phi}2 \right)} 
\ee
which is a simple traveling wave moving to the right with speed $v = |b|$, where $\phi\in\R$ is an arbitrary constant. The parameter $b<0$ corresponds to the unique eigenvalue of the Sturm--Liouville operator (stationary Schr\"odinger equation with potential $-u(x,0)$) 
\be
-f''(x)- u(x,0) f(x) = \frac{b}{4} f(x), \ \ \ f\in L^2(\R).
\label{SE}
\ee
If $f$ is the Jost solution, then the associated norming constant $\gamma$ and shift $\phi$ are given by $\gamma = \| f \|_2^{-1}$ and $\phi = 2\log \left(\frac{\gamma^2}{\sqrt{|b|}} \right)$. 

This particular solution can be written suggestively as 
\be
u(x,t) = 2\pa_x^2 \ln \le(1 + \frac { {\rm e}^{\sqrt {|b|} x - |b|^\frac 32 t {+\phi}} }{2\sqrt{|b|}}\ri).
\ee

Despite the equation being nonlinear, KdV admits $N$-soliton solutions that describe a superposition of the simple solitons introduced above. The $N$-soliton solution can be concisely described by the Kay-Moses \cite{KayMoses}  formula of Fredholm type:
\be
u(x,t) =&  2\pa_x^2 \ln \det \le[ \1_N + \mathbb G(x,t)\ri], \qquad \text{where}\cr
\mathbb G_{\ell m } (x,t) =& 
 \frac 
{ \sqrt{C_\ell C_m} {\rm e}^{\frac{\vartheta_\ell +\vartheta_m}2 }}{\sqrt{{|b_\ell |}} +\sqrt{{ |b_m| }}}, \ \  
\vartheta_\ell :=\sqrt{ { |b_\ell | }} x - {|b_\ell |}^\frac 3 2 t, \ \ \ell=1,\dots,N.
\label{KayMoses}
\ee
Here: $\1_N$ denotes the identity matrix of size $N$; $\mathbb G$ is the $N\times N$ matrix indicated above;
the parameters $b_\ell$ are arbitrary negative numbers, and; $C_\ell,\ \ell =1,\dots, N$, are arbitrary positive numbers. 
It is well known that for an $N$-soliton solution \eqref{KayMoses} the  spectrum  of  \eqref{SE} is 
$\{b_1,\dots,b_N\}$;  the constant  $C_\ell$  is called a {\it norming constant}  associated with $b_\ell$, $\ell=1,\dots,N$.

A different family  is the so-called {\it finite-gap} family of solutions \cite{ItsMatveev, Dubrovin-Novikov}, which can be written as 
\be
u(x,t) = 2\pa_x^2 \ln {\bs \tau}(x,t),
\ee
where the  tau--function ${\bs \tau}(x,t)$ for the 
finite-gap solutions is expressed in terms of the Riemann theta function associated to {underlying}  hyperelliptic Riemann surface. 
 (For the $N$--soliton solutions, the  tau--function is given  by the Kay-Moses determinant \eqref{KayMoses}.)
 
The starting point of this work is the obsrevation that $N$-soliton solutions can be obtained  from the finite-gap solutions by degenerating the hyperelliptic surface to a nodal curve of genus zero (first observed by Its and Matveev in  \cite{ItsMatveev}); the computation is contained essentially in the last chapter of  Mumford's   book  \cite{Mumford}, where  a determinantal formula of different type was derived (see also   \cite{Its}, \cite{Matveev} and \cite{Its82}, where the degeneration procedure was used and a determinant formula for the $N$-soliton solutions of the focusing NLS equation was obtained). In the recent work \cite{Gaillard} the Kay-Moses formula is recovered from the degeneration via the equivalence with the Wronskian formula of Matveev \cite{Matveev}.  

In this paper wepresent a generalization of the Kay-Moses formula to the case where the hyperelliptic curve is {\it partially} degenerated to a nodal curve of genus $1$. 
The core of the computation is, in fact, more general; it is based on a formula (presented in Appendix \ref{genusg}) for the limit of the Riemann Theta function (with appropriate characteristics) when the curve degenerates to a nodal curve whose resolution has an arbitrary genus $g$. 

A formula for the partial degeneration can also be found in 
the monograph \cite{belokolos} (Chapter 4, p. 138) in the context of the Nonlinear Schr\"{o}dinger equation).
This formula provides a different, not determinantal  description of  the degeneration (see Theorem  \ref{main} below),  which   also could be used to study the scattering of the NLS solitons on the cnoidal background. We now describe the setting of the problem.

Consider the real elliptic curve 
\be\label{el-curve}
w^2 = 4 z^3 - g_2 z -g_3 = 4(z-e_1)(z-e_2)(z-e_3), \ \ \ e_3<e_2<e_1, \ \ e_1 +e_2+e_3 = 0
\ee
with  half-periods 
\be
\varpi_1 :=\int_{e_3}^{e_2} \frac { \d z}{2 \sqrt{(z-e_3)(z-e_2)(z-e_1)_+}} \in \R_+
\\
{\varpi_3} :=\int_{e_1}^{e_2} \frac { \d z}{2 \sqrt{(z-e_3)(z-e_2)(z-e_1)}} \in i\R_-,
\ee
where the radical is chosen with branchcuts $[e_3,e_2]\cup[e_1,\infty)$ and with {the} determination such that it is in $i\R_-$ in the gap $[e_2,e_1]$. 
Then the stationary cnoidal wave solution (genus one finite-gap or one-phase nonlinear wave  solution) is given by 
\be
u(x,t) = 2\pa_x^2 \ln \bs \tau(x,t)=2\pa_x^2  \ln \left[  {\rm e}^{- \frac{\zeta({\varpi_3})}{{8}{\varpi_3}}x^2}  \theta_3\le(\frac {x}{4i{\varpi_3}},\tau \ri) \right],
\ee
where $\tau: = \varpi_1/{\varpi_3} \in i\R_+$,
and  $\theta_{1,2,3,4}(\beta,\tau)$ denote  the standard Jacobi elliptic theta functions.  

Note that the choice $e_1+e_2+e_3=0$ in \eqref{el-curve} is made without loss of generality. If $\tilde u(x,t)$ is a solution of \eqref{KdV-i} corresponding to the elliptic curve with $\wt e_1+\wt e_2+\wt e_3 = v$, then using the Galilean symmetry of \eqref{KdV-i} $\tilde u(x,t) = u(x-2v t, t)+ \tfrac{v}{3}$, where $u$ is a solution of \eqref{KdV-i} associated with an elliptic curve with branchpoints $e_1,e_2,e_3$  satisfying \eqref{el-curve}. In this way we can obtain any cnoidal travelling wave solution from the family of stationary cnoidal solutions.

Choose $L$ points $b_j\in(-\infty,e_3),   \ \ j=1,\dots L$ and $N-L$ points  
$b_j\in(e_2,e_1),   \ \ j=L+1,\dots N$. These points represent the  centers of  $O(\eps)$,  $\eps\to 0^+$,  fast  shrinking bands of some genus $N+1$ hyperelliptic Riemann surface $\Rscr_N(\eps)$.
If  $N$ is fixed, the rate of decay of the shrinking bands is not essential. (But in the case of soliton gases, where $N\to\infty$ (see below), $\eps$ is   linked with $N$ and the rate  of decay of the bands is important.)

Let $\Theta$ be the Riemann Theta function \cite{Fay} on $\Rscr_N(\eps)$:
\def\n{{\bf n}}
\be
\Theta\le(\bs X;  {\bs \Omega}\ri):= \sum_{{\bs \nu} \in \Z^{N+1} } {\rm e}^{i\pi {\bs \nu}^\transpose {\bs \Omega} {\bs \nu} + 2i\pi {\bs \nu}^\transpose {\bs X}}, \ \ \ {\bs X}\in \C^{N+1},
\ee
where ${\bs \Omega} = {\bs \Omega}(\eps)$ is the Riemann period matrix (our choice of  $\scr A_j$ and  $\scr B_j$ cycles is shown on Figure \ref{cycles}).  Let us set ${\bs X}=[\psi_1,\dots, \psi_N,\beta]^\transpose \in \C^{N+1}$.
Our main observation,  see Theorem \ref{main},   states that in the limit $\eps\to 0$ the Riemann Theta function 
\be\label{theta-factor}
\lim_{\eps\to 0}\Theta\le( {\bs X} - \frac 1 2 {\bs \Omega}(\eps) {\bs u};{\bs \Omega}(\eps) \ri) =
\det\le[
\1_N + \mathbb G\ri]
\theta_3\le(\beta - \mathcal A\ri),
\ee
  where 
 the shift  $ \mathcal A$ depends only on $b_1,\dots, b_N$,  
 and the $N\times N$ matrix $ \mathbb G$ 
 depends on $b_1,\dots, b_N$ and also on 
${\bs X}$.
   The limiting elliptic curve \eqref{el-curve} with $N$ pairs of identified points $b_1,\dots, b_N$ (on both sheets) will be  denoted by $ \Rscr_N(0)$.
  
The factorization  \eqref{theta-factor} represents the limiting Riemann Theta function as a product of Kay-Moses type determinant and the (shifted) Riemann Theta function $\theta_3$ on the residual elliptic background $ \Rscr_N(0)$.  It is a direct generalization of Mumford's approach to the case when all but one bands of the Riemann surface are shrinking as $\eps\to 0$ and, thus, in the limit, one obtains an $N$-soliton solution to the KdV on the residual (elliptic) background. In Appendix A this result is generalized to an arbitrary hyperelliptic residual background.

Our next main result, Theorem \ref{solitons!}, states that
\be
u(x,t)= 2\pa_x^2 \ln {\bs \tau}(x,t)
\quad \text{with} \quad
{\bs \tau}(x,t):= {\rm e}^{- \frac{\zeta({\varpi_3})}{{8}{\varpi_3}}x^2} \det \le[\1_N+\mathbb G(x,t)\ri] \theta_3\le(\frac x{4i{\varpi_3}}-\mathcal A\ri)
\ee
is a solution to  the KdV \eqref{KdV-i},
where $\zeta$ denotes the Weierstrass zeta function on the elliptic curve \eqref{el-curve}  and  the $x,t$ dependence of the vector ${\bs X}$, which is part of   $\mathbb G$,  is defined in terms of the limiting  quasi-momentum and quasi-energy  meromorphic differentials on $\Rscr_N(\eps)$ as $\eps\to 0$, see Theorem \ref{solitons!}  for details.
Thus,
\be\label{sol-gen-1}
u(x,t)=2\pa_x^2 \ln \det \le[\1_N+\mathbb G\ri] + 2\pa_x^2 \ln \theta_3\le(\frac {x}{4i{\varpi_3}}-\mathcal A\ri) - \frac{\zeta({\varpi_3})}{{2}{\varpi_3}}, 
\ee
i.e., up to a constant, solutions of KdV describing 
the $N$ solitons on the elliptic background 
  can be represented as a sum of  the Kay-Moses type determinant ``tweaked by the elliptic background" and the elliptic background  solution  $2\pa_x^2 \ln \theta_3(y)$ ``shifted by the $N$ solitons".

\begin{figure}
	\begin{tikzpicture}[scale=2]

	\fill[black!05!white] (3.5,0.5) rectangle (-4.5,-0.5);
		\coordinate (e1) at (1,0); \node at (e1)[below] {$e_1$};
		\coordinate (e2) at (-1,0); \node at (e2) [below] {$e_2$};
		\coordinate (e3) at (-2,0) ; \node at (e3) [below] {$e_3$};
		\draw[thick] (e1) to (2.5,0);
		\draw[thick] (e3) to (e2);
		\draw[dashed ] (2,0) to (3,0);
		\foreach \b [count=\bk] in {-4, -3.7, -3.05,-2.8,-2.4,-2.2} {
			\draw [red, fill=red] (\b,0) circle [radius=0.02] node [below] {$b_{\bk}$} ;
			\draw [thin, red] ($ (\b,0) - (0.07, 0) $)  -- ($ (\b,0) + ( 0.07,0) $) ;
		}
		\foreach \c [count=\ck from 7] in {-0.8, -0.6, -0.3, 0.1,  0.75} {  
		\draw [blue, fill=blue] (\c,0) circle [radius=0.02] node [below] {$b_{\ck}$} ;
			\draw [thin, blue] ($ (\c,0) - (0.07, 0) $)  -- ($ (\c,0) + ( 0.07,0) $) ;
		}
		
	\end{tikzpicture}
	\caption{The position of hot (red), $j=1,\dots, L$, and cool (blue), $j=L+1,\dots,N$, points $b_j$ of discrete spectra, relative to the spectral bands $[e_3,e_2]\cup [e_1,\infty)$ of the elliptic background. 
		These can be viewed as the degeneration of small bands---centered at each discrete spectral point---on a higher genus surface.}
	\label{spectrum}
\end{figure}
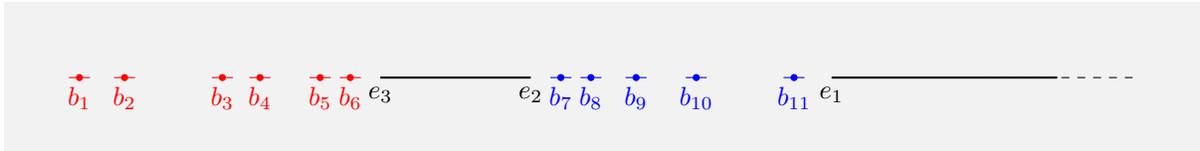

\begin{figure}
\centerline{
\includegraphics[width=0.3\textwidth]{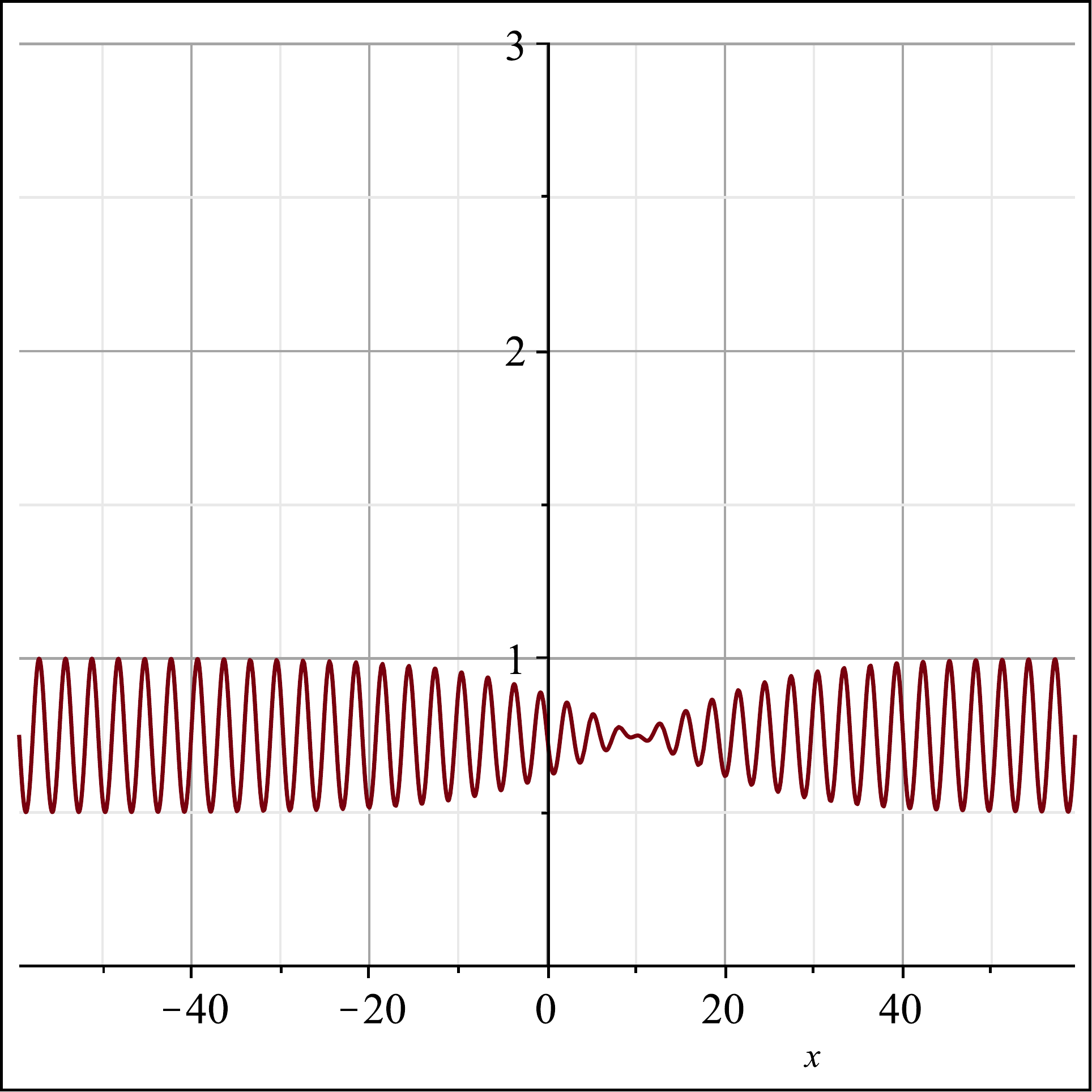}\ \ \ 
\includegraphics[width=0.3\textwidth]{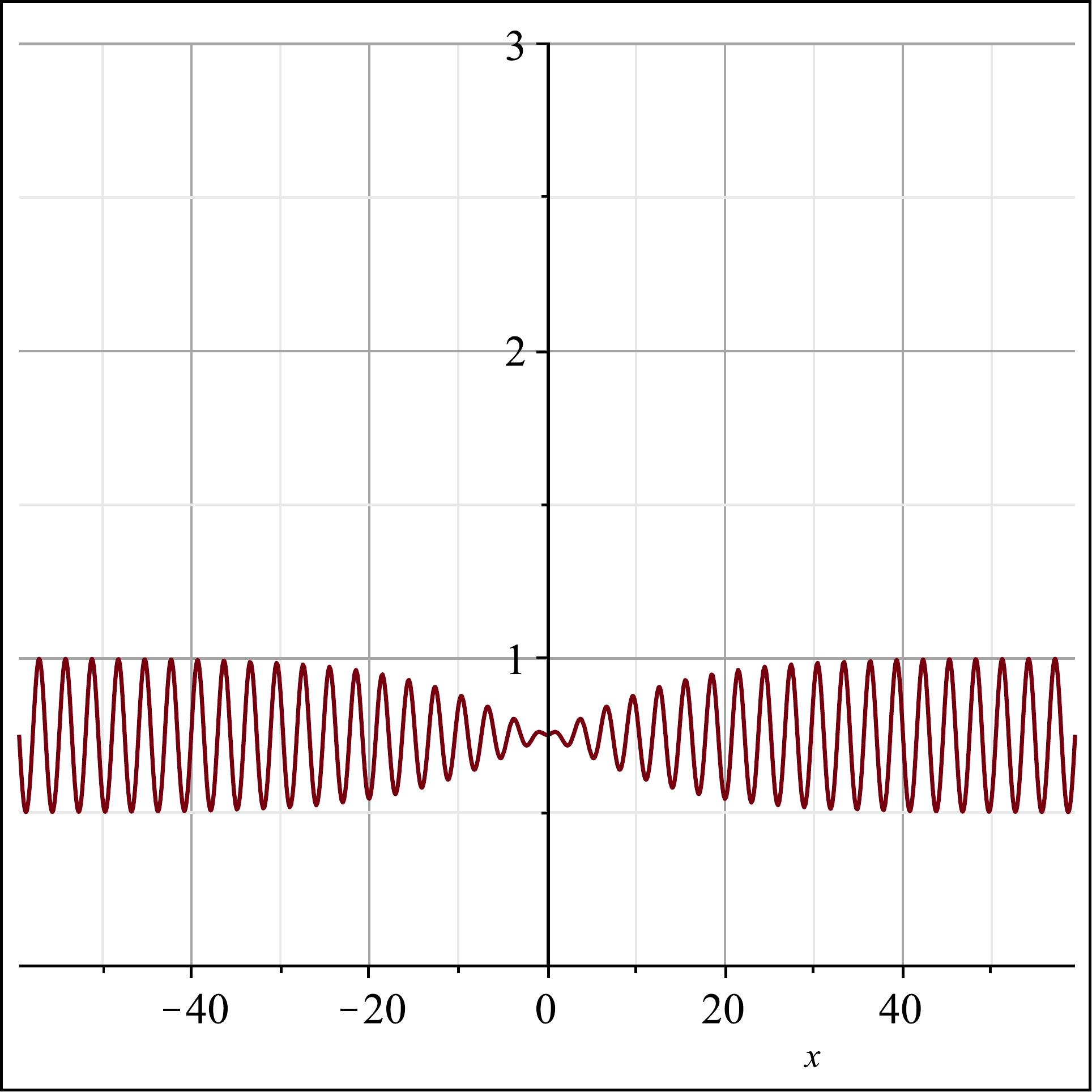}\ \ \ 
\includegraphics[width=0.3\textwidth]{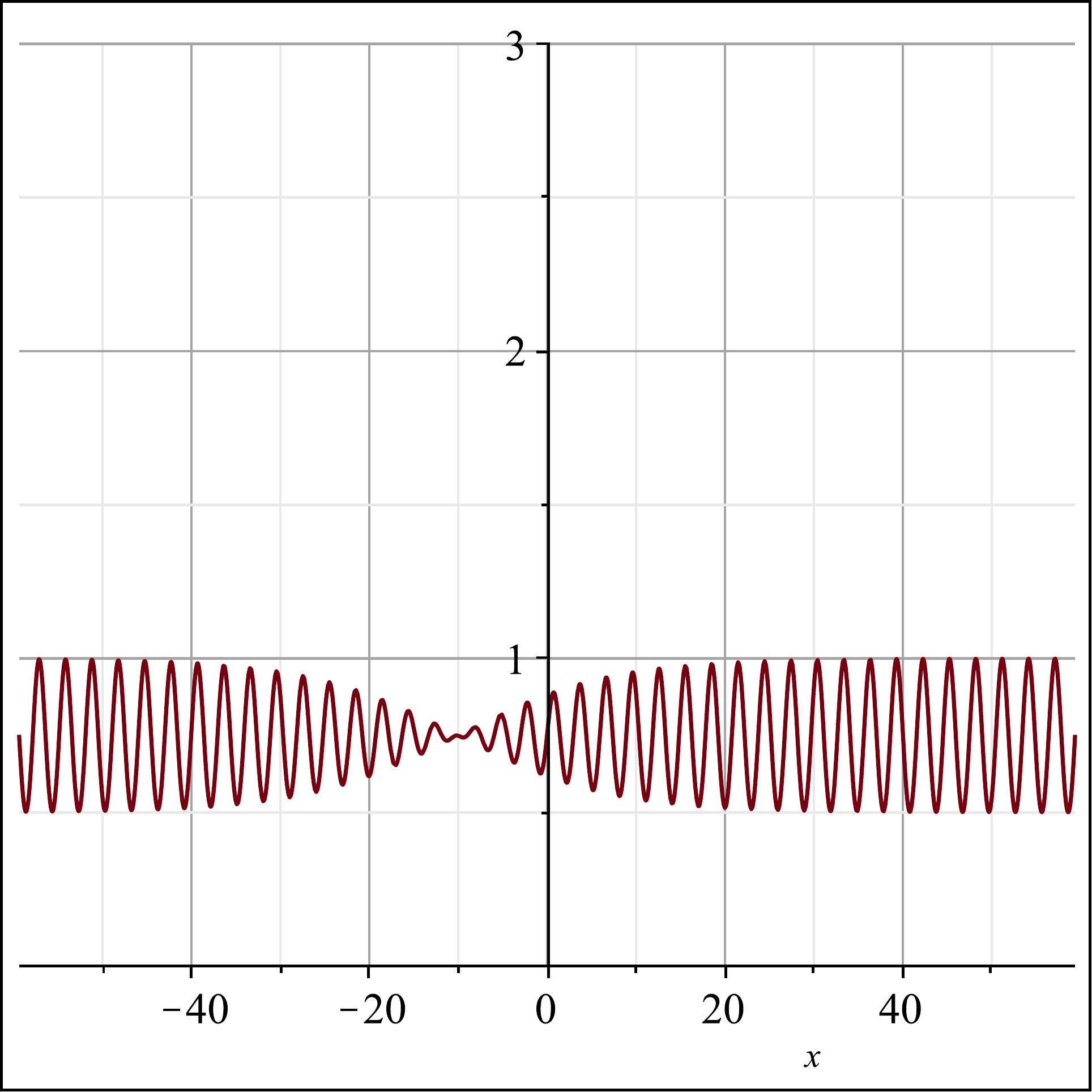}
}
\caption{Plot of a dim retrograde soliton at three different times; the parameters of the elliptic curve are as in  figure~\ref{groupvelfig} and $\beta = 0.24 + \frac {\tau} 2$ ($c\simeq 1.50356 $). The group velocity calculated with formula \eqref{TrackerVelocity} is $V \simeq -8.99139$. In the left/right pane the time is set to $\mp 10/|V|$ to see that the disturbance has travelled exactly $10$ units (and towards the left).}
\label{dim}
\end{figure}

\begin{figure}
\centerline{
\includegraphics[width=0.3\textwidth]{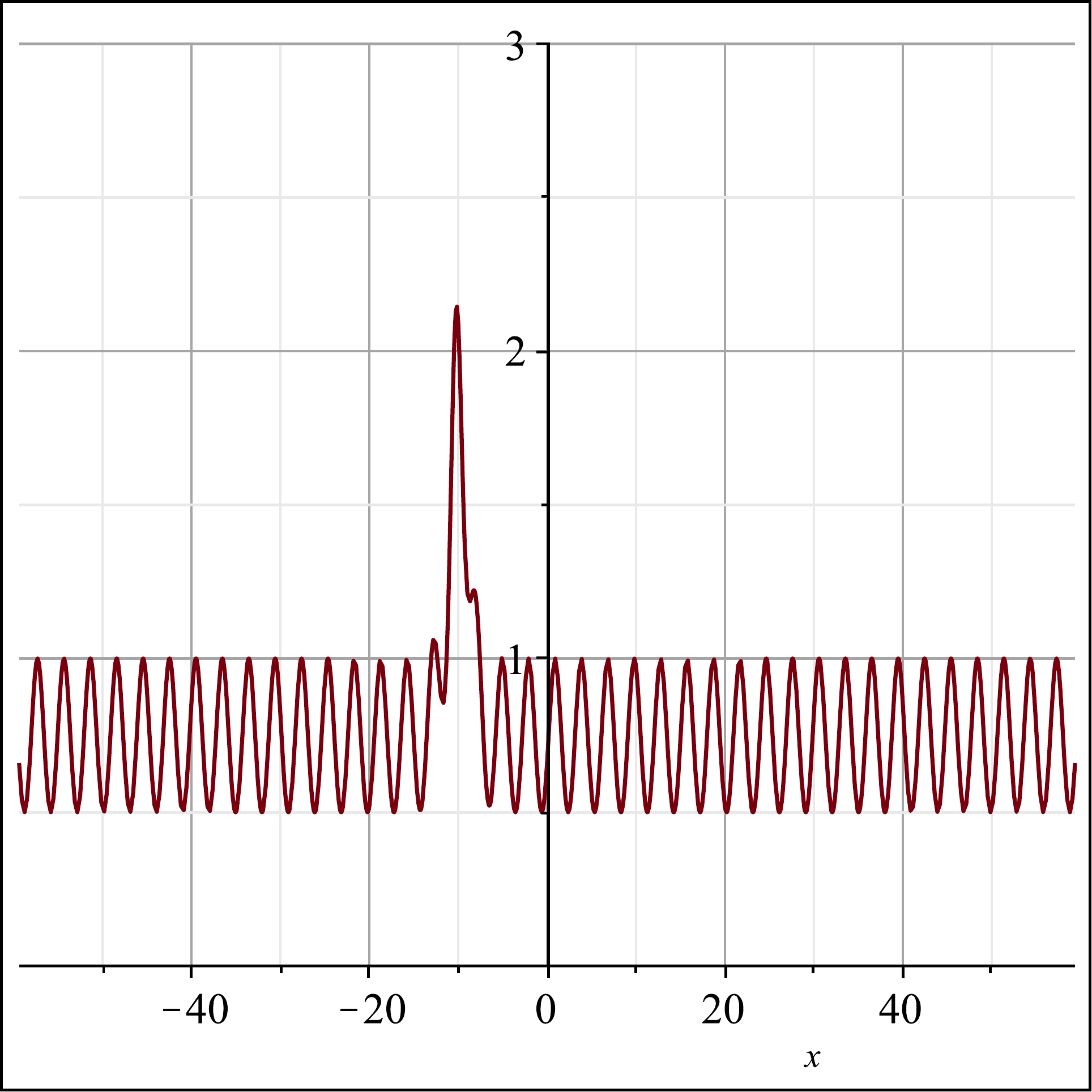}\ \ \ 
\includegraphics[width=0.3\textwidth]{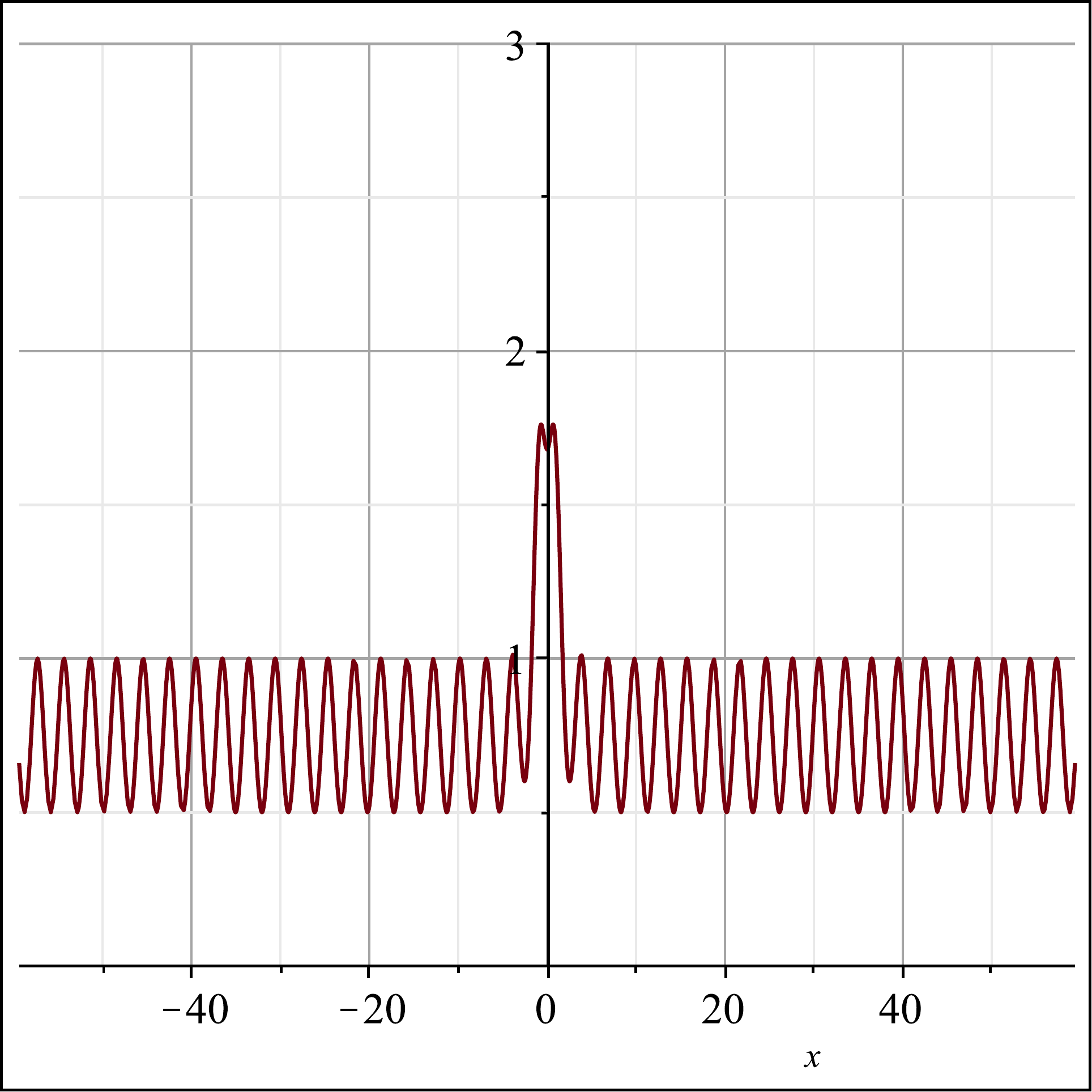}\ \ \ 
\includegraphics[width=0.3\textwidth]{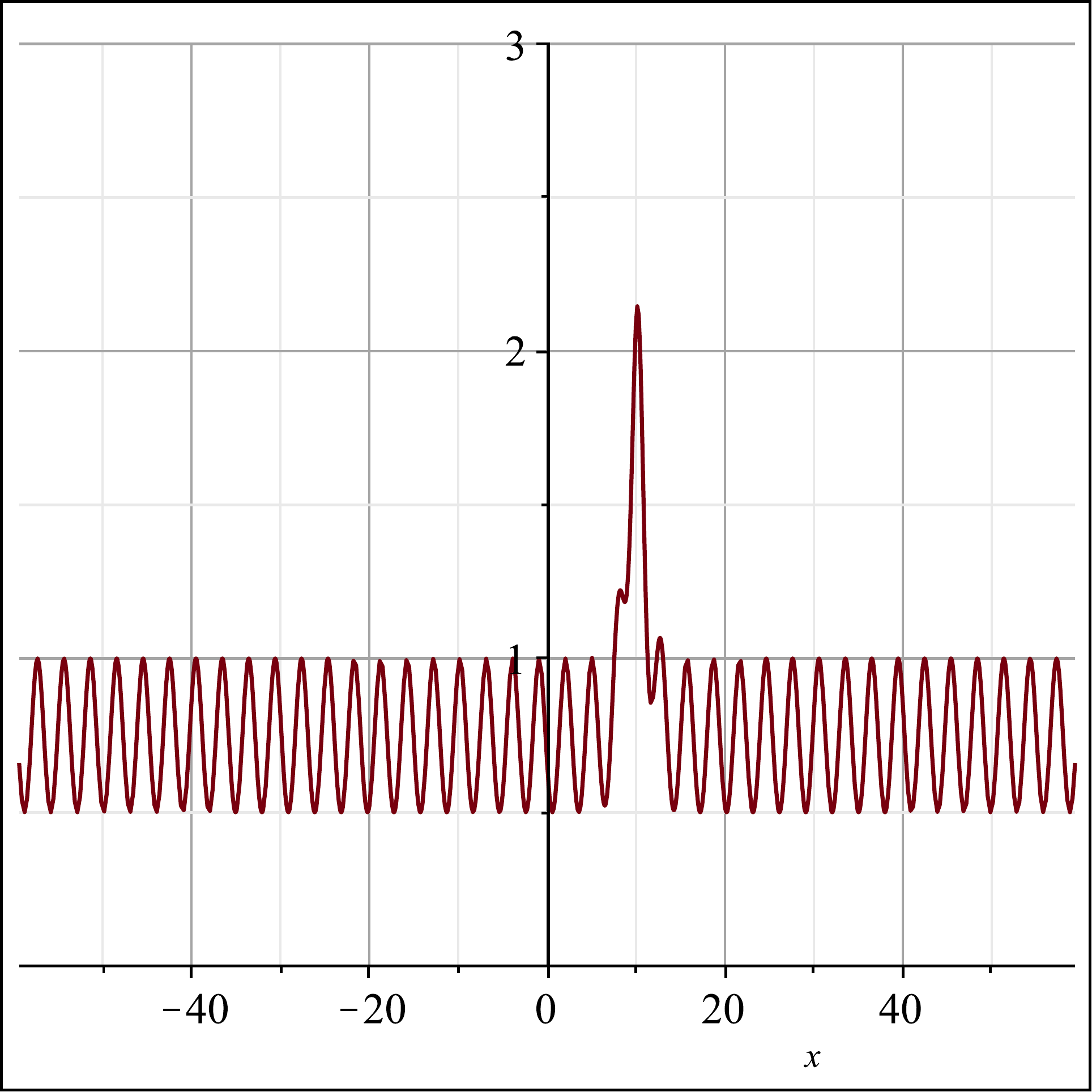}
}
\caption{Plot of a bright forward soliton at three different times; the parameters of the elliptic curve are as in figure~\ref{groupvelfig} and $\beta =  0.30$ ($b\simeq -5.3595$). The group velocity calculated with formula \eqref{TrackerVelocity} is $V \simeq 6.8273$. In the left/right pane the time is set to $\pm10/|V|$ to see that the disturbance has travelled exactly ten units.}
\label{bright}
\end{figure}

Proofs of Theorems \ref{main} and \ref{solitons!}, including  the appropriate notations from algebraic geometry, are the subject of Section \ref{sect-proofs}. Important steps in these proofs are the calculation of the Riemann period matrix ${\bs \Omega}$, the normalized holomorphic differentials and the quasi-momentum and quasi-energy differentials in the limit $\eps\to 0$. 
Using results of  Section \ref{sect-proofs}, we then calculate 
the velocity of a single soliton on the elliptic background (Section \ref{sect-speed}) and the phase shift of two interacting solitons on the elliptic background (Section \ref{scattering}). Here we want to mention that the solitons corresponding to $b_j<e_3$, see Figure \ref{spectrum}, appear to have larger than the cnoidal wave amplitude and positive speed (bright-and-forward or simply hot),  see
Figure \ref{bright}.
On the other hand, the solitons corresponding to $b_j\in(e_2,e_3)$ appear to have smaller than the cnoidal wave amplitude and negative  speed (dim-and-retrograde or simply cool).  The   smaller than the cnoidal wave amplitude 
of the soliton can be identified with the dip in the cnoidal wave oscillations 
clearly visible on 
Figure \ref{dim}; 
	for some early works about
	  solitons on the elliptic background see,  for example, \cite{Wahlquist}, \cite{KuMi}.

Dim (and retrograde) solitons, also mentioned  as solitary disturbances, have  {\it negative group velocity}, i.e., they are moving from the right to the left, which runs contrary to the common understanding that the solitons for the KdV equation \eqref{KdV-i} are moving left to right. 
	However, the motion in the opposite direction is the result  of the interaction between the soliton and the background. A similar phenomenon happens in  KdV soliton gases on zero background,   where the faster (and taller) solitons are constantly pushing back the slower (and smaller ) solitons due to the phase shift of their pair-wise interaction, so that the effective velocity of sufficiently small solitons is negative, see \cite{CERT}.  Numerical observation of KdV solitons moving in the negative direction was first reported in \cite{Pelin-F}.

In Section \ref{sect-gas} we use the results described  above  to derive the main equations for a KdV soliton gas on the elliptic background, such as the nonlinear dispersion relations (NDR) and the equation of state.  The concept of a soliton gas, which can be traced back to some ideas of V. Zakharov \cite{Zakh-71} and S. Venakides \cite{Ven-89}, was  formulated by G. El in \cite{El2003}. 
A soliton gas can be considered as a large $N\to\infty$ limit of   an ensemble of $N$ solitons, viewed as particles with 2-particle interactions.  Alternatively, it can be viewed as a specific  
(thermodynamic) limit of genus $N$ finite-gap solutions when $N\to\infty$ and simultaneously the size of all (or all but finitely many) bands go to zero exponentially  fast in $N$. 
Thus, we are interested in the large $N$ limit of various quantities associated with a hyperelliptic Riemann surface $\Rscr_N(e^{-\nu N})$, where $\nu>0$. In this setting the NDR become simply the thermodynamic limit of the Riemann bilinear identities on $\Rscr_N(e^{-\nu N})$, which  involve the quasi-momentum and quasi-energy meromorphic differentials on one side and the normalized holomorphic differentials  on the other.  
Many details  about soliton gases for the NLS and KdV equation can be found in \cite{ElTo2020,El2003}, see also \cite{El-Rev}, \cite{TW2022}.  In particular,  the thermodynamic limits of the NDR were derived  in \cite{ElTo2020}  for the focusing NLS  soliton (all bands are shrinking) and breather (all but one band are shrinking) gases. In both cases, the genus of the residual (degenerate) surface was zero.  In Section   \ref{sect-gas} we make the next step in this direction by deriving  the NDR and the equation of state for a KdV gas on the genus one background.  In light of Appendix \ref{genusg}, it is quite  clear that one can use the same technique to derive the NDR for a KdV gas on backgrounds of any finite genus.
We want to mention that the error estimate of the limiting NDR is   outside the scope of this paper, although some partial results related to this issue {in the context of the NLS soliton gas} can be found in  \cite{TW2022}. 

In the Appendices \ref{genusg} and \ref{sect-aver} we respectively state and sketch the proof of Theorem \ref{main} for any genus $g\geq 1$ background (residual) Riemann surface, as well as prove that any solution described by Theorem \ref{solitons!}  converges to the backgound elliptic solution with respect to some natural  probability measure uniformly on compact subsets of $(x,t)$.

The following theorem summarizes the main results of Sections \ref{sect-proofs} - \ref{scattering}. Its proof follows from Theorems \ref{main}, \ref{solitons!}, \ref{scatteringthm}.
Proof of \eqref{Total-shift} requires repeated use of Theorem  \ref{scatteringthm}.

Fix $b_1 < b_2 < \dots < L \in (-\infty, e_3)$ and $b_{L+1} < \dots < b_N \in (e_2, e_3)$. 
Let $\wp(s)$ denote the Weierstass function on the elliptic curve \eqref{el-curve}. Define implicitly $\beta_k$ satisfying  $b_k =\wp(2{\varpi_3}\beta_k),$  $k=1,\dots,N, $ and 
$\Re(\beta_k)\in [0,1), \  \Im(\beta_k)\in \{ 0,\Im\tau/2 \}$.  Denote by $\beta_k^\star:= 1- \beta_k + \chi\tau$ the second pre-image of each $b_k$ {in the fundamental rectangle}, where $\chi=1$ if $\Im \beta= \Im \tau/2$ and $\chi=0$ if $\Im \beta=0$.

\begin{figure}
\centerline{
\includegraphics[width=0.3\textwidth]{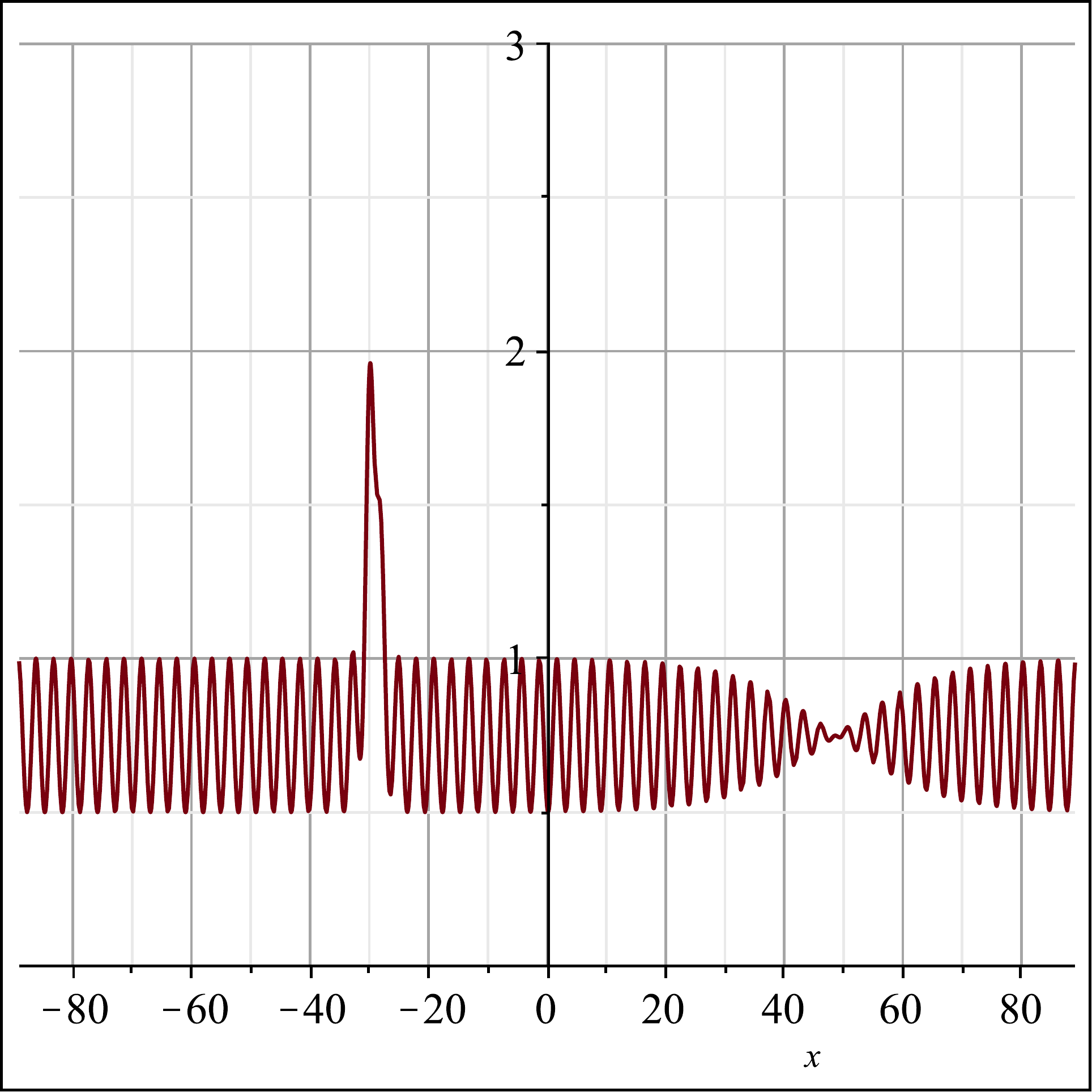}\ \ \ 
\includegraphics[width=0.3\textwidth]{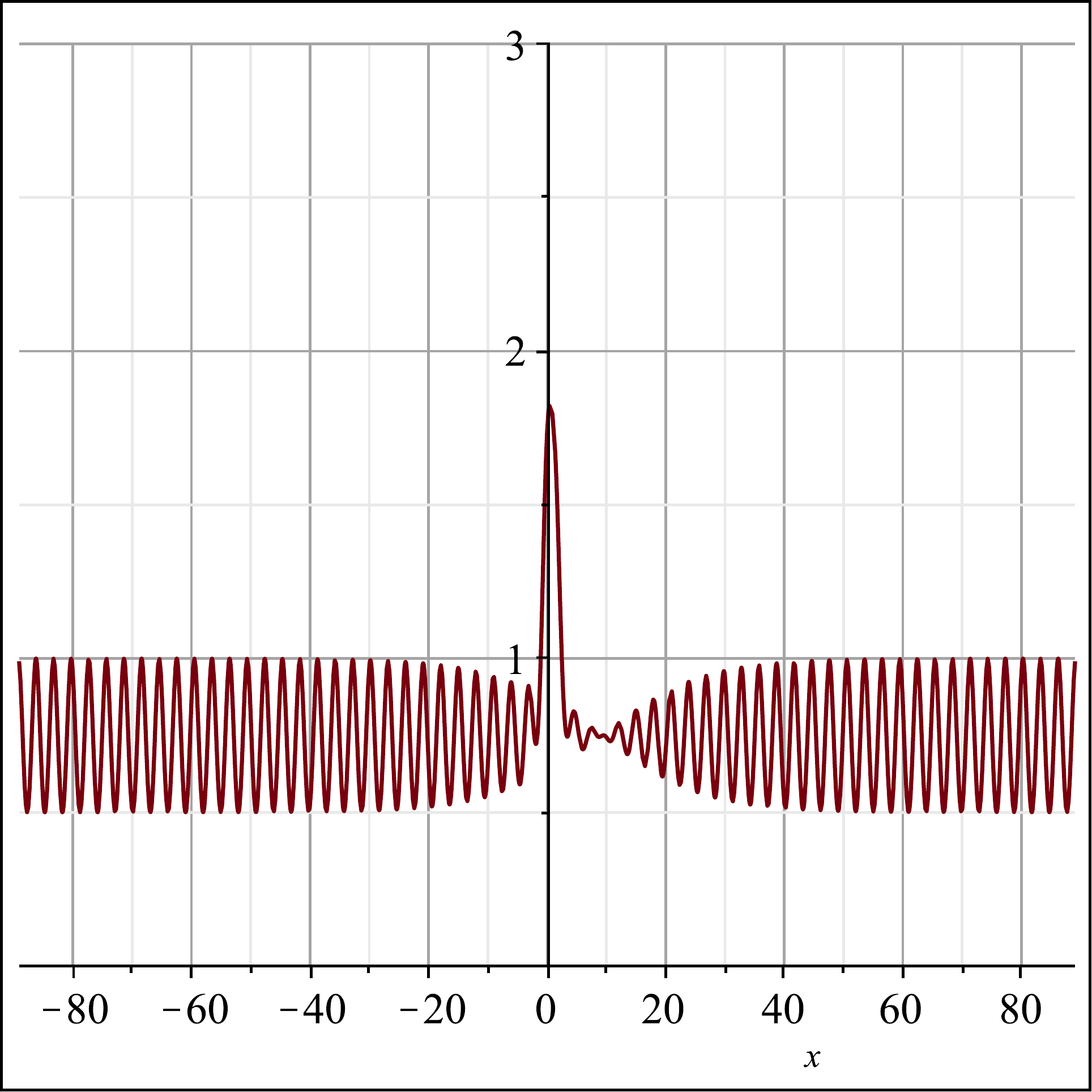}\ \ \ 
\includegraphics[width=0.3\textwidth]{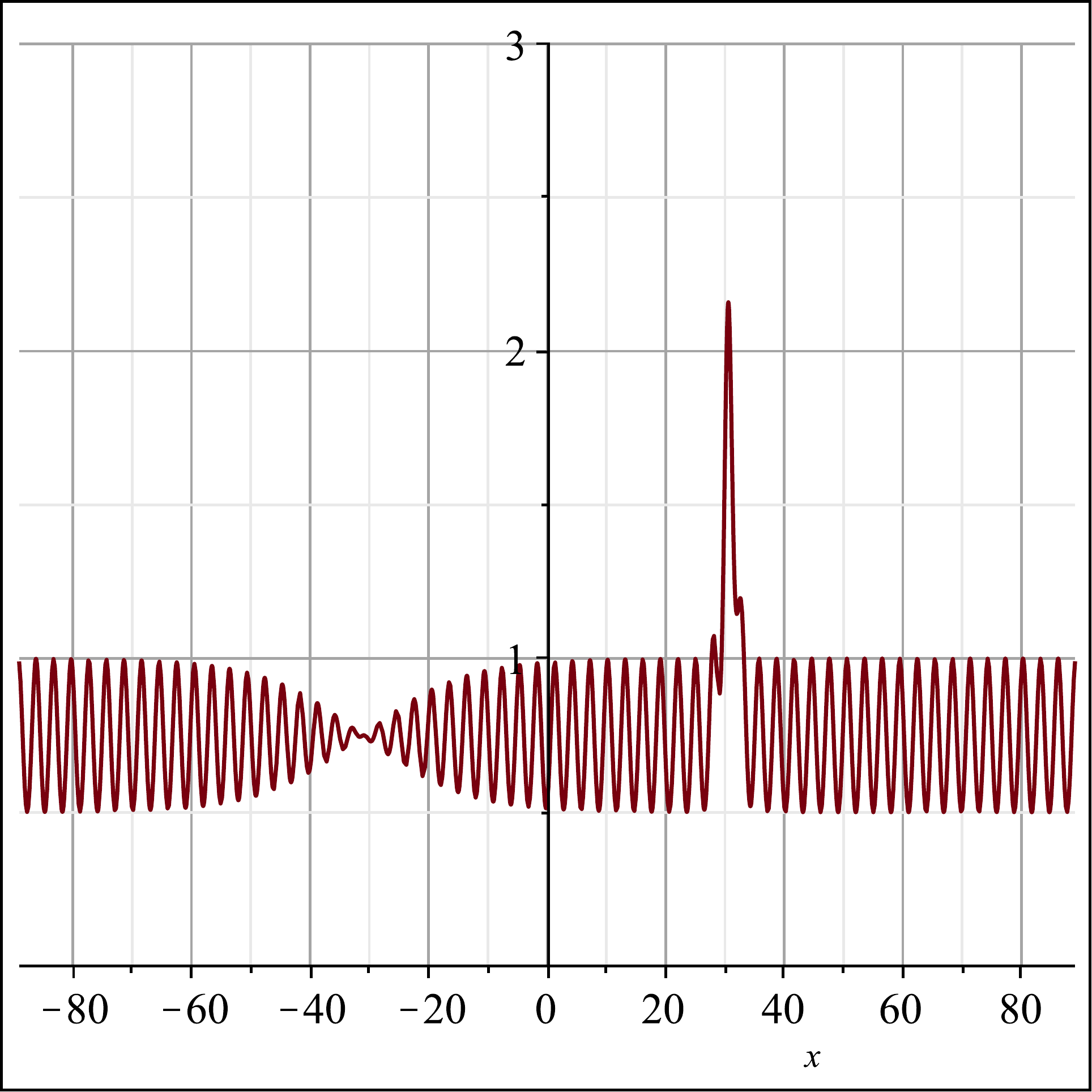}
}
\caption{Plot of the two--soliton solution with both a bright and a dim soliton at times $-30/v_{hot},0,30/v_{hot}$. The parameters of the two solitons are the same as in Fig. \ref{dim} and Fig. \ref{bright}. As expected, the dim soliton travels to the left and the bright one to the right with their own group velocities. }
\label{dimbright}
\end{figure}

\bt 
\label{thm:1.1}
{\bf [1]} 
The solution of the KdV equation \eqref{KdV-i} with
$N$ solitons on a cnoidal background satisfying \eqref{el-curve} is given by 
\begin{gather}
	u(x,t)=2\pa_x^2 \ln \det \le[\1_N+\mathbb G\ri] + 2\pa_x^2 \ln \theta_3\le(\frac {x-x_0}{4i{\varpi_3}}-\mathcal A\ri) - \frac{\zeta({\varpi_3})}{{2}{\varpi_3}}, 
	 \cr
\text{{where the background shift }}\quad  \mathcal A:= \frac 1 2 \sum_{j=1}^N (\beta_j - \beta_j^\star), \cr
\mathbb G_{\ell, m}:=  
\frac {
\ds 
\theta_3\le(\beta_\ell - \beta^\star_m +\frac {x-x_0}{4i{\varpi_3}}   -  \mathcal A\ri)}{ 
\ds
\theta_1\le( \beta_\ell - \beta^\star_m \ri) \theta_3\le(\frac {x-x_0}{4i{\varpi_3}}  -  \mathcal A\ri) }
\sqrt{C_\ell C_m} {\rm e}^{i\pi(\psi_\ell+\psi_m)},\quad \ell,m=1,\dots,N, \cr
\psi_j(x,t):= (x-x_j^{(0)})\frac{P_j}{2\pi} + t \frac {E_j}{2\pi}, \qquad
E_j:= - \frac 1 2\wp'(2{\varpi_3}\beta_j), \qquad P_j:= 
{ \frac 1{2{\varpi_3}} \frac{\theta_1'\le( \beta ;\tau\ri)}{\theta_1\le( \beta ;\tau\ri)} \Bigg|_{\beta=\beta_j^\star}^{\beta=\beta_j}}
\end{gather}
and the norming constants $C_j$ are  positive numbers given explicitly in Theorem \ref{main} and $x_j^{(0)}, x_0\in\R$ are arbitrary shifts.
\\[10pt]
{\bf [2]} The points $\beta_j\in (0,\frac 1 2), \ \ j=1,\dots, L$ correspond to right-propagating solitons (solitary disturbances)  whereas the points $\beta_j\in (0,\frac 1 2) + \frac \tau 2, \ \ j\geq L+1$  correspond to left propagating solitons (solitary disturbances). 
\\[10pt]
{\bf [3]} The profile of such solutions for $t\to \pm \infty$ consists of a cnoidal stationary background (with period $X=4i{\varpi_3}\in \R^+$) modulated by solitons (solitary disturbances)  that are localized around the lines $x_j = V_j t + \Phi_j^{(\pm)}$. 
Each $V_j = -\frac {E_j}{P_j}$ gives the modified velocity of the solitons on the elliptic background; the order of velocities is preserved, i.e., $V_1 > V_2 > \dots$.
The phase shifts $\Phi_j^{(\pm)}$ depend on the norming constants, but their averaged difference ({the averaged total shift of the $j$th soliton}) does not: 
\be\label{Total-shift}
\le\langle\Phi^{(+)}_j \ri\rangle- \le\langle\Phi_j^{(-)}\ri\rangle = &
 \frac 2 {|P_j|}\sum_{k> j} \ln \left| \frac { 
 \theta_1(\beta_j-\beta^\star_{{k}})
}{
\theta_1(\beta_{{j}}-\beta_k)
} \right|
 - \frac 2 {|P_j|}\sum_{k< j} \ln \left| \frac { 
 \theta_1(\beta_j-\beta^\star_{{k}})
}{
\theta_1(\beta_{{j}}-\beta_k)
} \right|
\ee
where the average is over the period of the cnoidal wave (see description in Section \ref{scattering} and Figures \ref{dim}, \ref{bright}, \ref{dimbright}).

\et

\br
Part \textbf{[3]} in Theorem~\ref{thm:1.1} shows that solitons on the elliptic background affect a phase shift in the cnoidal background. A consequence of this fact is that the solitary disturbance is not localized. For $N=1$ this can be directly verified by computing the asymptotic behavior of the solitonic disturbance $2\partial_x^2 \ln (1 + \mathbb{G})$ as $x \to \pm \infty$. One finds that 
\[
	2\partial_x^2 \ln (1 + \mathbb{G}) \sim \begin{cases} 0 & x\to +\infty \\
	2\partial_x^2 \ln \theta_3\le(\beta_1 - \beta^\star_1 +\frac {x - x_0}{4i{\varpi_3}}   -  \mathcal A\ri) - 
	2\partial_x^2 \ln \theta_3\le(\ \frac {x-x_0}{4i{\varpi_3}}   -  \mathcal A\ri)
	& x \to - \infty
	\end{cases}
\]
\er


\textbf{Acknowledgements.}
The authors would like to thank the Isaac Newton Institute for Mathematical Sciences for support and hospitality during the programme {\it Dispersive hydrodynamics: mathematics, simulation and experiments, with applications in nonlinear waves}  when work on this paper was undertaken. This work was supported by EPSRC Grant Number EP/R014604/1.

M.B. is supported in part by the Natural Sciences and Engineering Research Council of Canada (NSERC) grant RGPIN/261229-2016.
A.T.  is supported in part by the National science Foundation grant DMS-2009647 and 
by a grant from the Simons Foundation.
R.J. is supported in part by Simons Foundation grant 853620.
The authors are grateful to G. El and A. Veselov for useful comments.

\section{Generalities about nodal hyperelliptic and elliptic curves}
\label{sect-proofs}
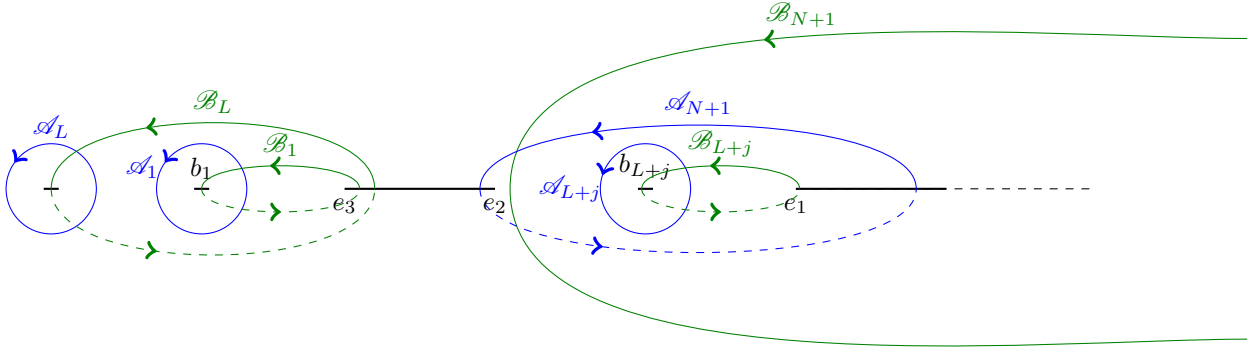
\begin{figure}
\begin{tikzpicture}[scale=2]
\draw[dashed ] (1,0) to (3,0);
\draw[thick] (1,0) to (2,0);
\node [below] at (1,0) {$e_1$};
\node [below] at (-1,0) {$e_2$};
\node [below] at (-2,0) {$e_3$};

\node [above] at (-2.95,0) {$b_1$};
\node [above] at (0,0) {$b_{L+j}$};

\draw[thick] (-2,0) to (-1,0);
\draw [thick] (-3,0) to (-2.9,0);
\draw [thick] (-4,0) to (-3.9,0);
\draw [thick] (-0.05,0) to (0.05,0);
\draw [blue,  postaction={decorate,decoration={markings,mark=at position 0.4 with {\arrow[line width=1.5pt]{>}}}}]  (-2.95,0) circle [radius=0.3];
\draw [blue, postaction={decorate,decoration={markings,mark=at position 0.4 with {\arrow[line width=1.5pt]{>}}}}] (-3.95,0) circle [radius=0.3];
\draw [blue,  postaction={decorate,decoration={markings,mark=at position 0.45 with {\arrow[line width=1.5pt]{>}}}}] (0.,0) circle [radius=0.3];

\node[blue] at ($(-2.95, 0) + (160:0.42) $)  {$\scr A_{1}$};
\node[blue] at ($(-3.95, 0) + (90:0.41) $)  {$\scr A_{L}$};
\node[blue] at ($(0, 0) + (180:0.5) $)  {$\scr A_{L+j}$};

\draw [green!50!black, postaction={decorate,decoration={markings,mark=at position 0.3 with {\arrow[line width=1.5pt]{>}}}}]  (4,1) 
to [out=180, in=90, looseness=0.7] node[pos=0.5,above]{$\scr B_{N+1}$} (-0.9,0)
to [out=-90, in=180, looseness=0.7]  (4,-1);
\draw [blue , postaction={decorate,decoration={markings,mark=at position 0.3 with {\arrow[line width=1.5pt]{<}}}}]  (-1.1,0) to [out=90, in=90, looseness=0.5] node[pos=0.5, above] {$\scr A_{N+1}$} (1.8,0); 
\draw [blue , dashed, postaction={decorate,decoration={markings,mark=at position 0.3 with {\arrow[line width=1.5pt]{>}}}}]  (-1.1,0) to [out=-90, in=-90, looseness=0.5]  (1.8,0); 

\draw [green!50!black , postaction={decorate,decoration={markings,mark=at position 0.5 with {\arrow[line width=1.5pt]{<}}}}]  (-2.95,0) to [out=90, in=90, looseness=0.5] node[pos=0.5, above] {$\scr B_{1}$} (-1.9,0); 
\draw [green!50!black , dashed, postaction={decorate,decoration={markings,mark=at position 0.5 with {\arrow[line width=1.5pt]{>}}}}]  (-2.95,0) to [out=-90, in=-90, looseness=0.5] (-1.9,0);

\draw [green!50!black , postaction={decorate,decoration={markings,mark=at position 0.5 with {\arrow[line width=1.5pt]{<}}}}]  (-0.025,0) to [out=90, in=90, looseness=0.5] node[pos=0.5, above] {$\scr B_{L+j}$} (1.025,0); 
\draw [green!50!black ,dashed, postaction={decorate,decoration={markings,mark=at position 0.5 with {\arrow[line width=1.5pt]{>}}}}]  (-0.025,0) to [out=-90, in=-90, looseness=0.5]  (1.025,0);


\draw [green!50!black , postaction={decorate,decoration={markings,mark=at position 0.35 with {\arrow[line width=1.5pt]{<}}}}]  (-3.95,0) to [out=90, in=90, looseness=0.7] node[pos=0.5, above] {$\scr B_{L}$} (-1.8,0); 
\draw [green!50!black ,dashed, postaction={decorate,decoration={markings,mark=at position 0.35 with {\arrow[line width=1.5pt]{>}}}}]  (-3.95,0) to [out=-90, in=-90, looseness=0.7]  (-1.8,0); 
\end{tikzpicture}

\caption{The choice of cycles around the main bands and around the shrinking bands near the hot (far left) and cool (middle gap) solitons.}
\label{cycles}
\end{figure}

We consider the degenerating Riemann surface described by the affine equation 
\be
\label{degRS}
Y^2 =4(z-e_3)(z-e_2)(z-e_1) \prod_{j=1}^N  (z-b_j+\varepsilon)(z-b_j+ \varepsilon)
\ee
where  (see Fig. \ref{cycles})
\be
{b_1<b_2<\dots < b_L} < e_3 < e_2< { b_{L+1} <\dots < b_N} < e_1; \ \ \  
\ee
Let us introduce the elliptic curve  (in Weierstrass form), see, for example, \cite{Hancock},  \cite{Akhiezer},  or any standard reference:
\be
\label{elliptic}
w^2 = 4(z-e_3)(z-e_2)(z-e_1).
\ee
The determination of $W(z):={2} \sqrt{(z-e_3)(z-e_2)(z-e_1)}$ is chosen with the branch-cuts along $[e_3,e_2]\cup [e_1,\infty)$ so that
\be
W(z_+)\in \R_-, z\in [e_1,\infty],\ \ W(z_+) \in \R_+ ,\ z\in [e_3,e_2],\\
W(z)\in i\R_-, \ z\in [e_2,e_1],\ \ \ W(z)\in i\R_+, \ \ z \in (-\infty,e_3].
\label{2.5}
\ee

We denote the two half periods ${\varpi_3},\varpi_1$ and the modular parameter $\tau$ as follows
\be
\label{wlog}
\int_{e_3}^{e_2} \frac { \d z}{2 \sqrt{(z-e_3)(z-e_2)(z-e_1)_+}} =\varpi_1 \in \R_+
\\
\int_{e_1}^{e_2} \frac { \d z}{2 \sqrt{(z-e_3)(z-e_2)(z-e_1)}} ={\varpi_3}\in i\R_-
\\
\tau =\frac {\varpi_1}{{\varpi_3}} \in i \R_+.
\ee

With these understandings we
define the Abel map by
\be
\beta = \int_{-\infty}^z \frac {\d z}{{2}{\varpi_3} W(z)}.
\ee
In particular the images of $e_3,e_2,e_1$ are $\frac \tau 2, \frac {\tau+1}2, \frac 1 2$, respectively. 
 We recall the definitions and the main properties of the fundamental Weierstrass $\zeta$ and $\wp$ functions:
\begin{gather}
\zeta(s) := \frac 1 s + \sum'_{n,m}\le(\frac 1{s-2m{\varpi_3}-2n\varpi_1} +\frac 1{2m{\varpi_3}+2n\varpi_1}+\frac s{(2m{\varpi_3}+2n\varpi_1)^2}
\ri)
\\
\zeta(s+2{\varpi_3}) = \zeta(s) + 2\zeta \le({\varpi_3}\ri),
\qquad 
\zeta(s+2\varpi_1) =\zeta(s) + 2 \zeta\le(\varpi_1\ri)
\label{periodzeta}
\\
\wp(s):= -\zeta'(s)=\frac 1 {s^2} + \sum'_{n,m}\le(\frac 1{(s-2m{\varpi_3}-2n\varpi_1)^2} - \frac 1{(2m{\varpi_3}+2n\varpi_1)^2}\ri),\ \\ 
\wp({\varpi_1}) = e_1; \qquad \wp({\varpi_3}+\varpi_1) = e_2; \qquad \wp(\varpi_3) = e_3\\
(\wp')^2 = 4\le(\wp-e_1\ri)\le(\wp-e_2\ri)\le(\wp-e_3\ri). \label{WP}
\end{gather}
We will also need the Jacobi theta functions with the specific normalization below, which  differs from the one in DLMF \cite[\href{http://dlmf.nist.gov/20.2.i}{20.2(i)}]{DLMF}, by a factor of $\pi$; for example, for us here $\theta_3$ is $1$--periodic instead of $\pi$ periodic.
\be
\theta(\beta)&= \theta_3(\beta;\tau) := \sum_{n\in \Z}{\rm e}^{i\pi n^2\tau + 2i\pi n \beta}
\\
\theta_1(\beta) &= \theta_1(\beta;\tau) := \sum_{n\in \Z}
{\rm e}^{i\pi \le(n-\frac 12 \ri)^2\tau + 2i\pi \le(n-\frac 1 2 \ri)\le( \beta-\frac 1 2\ri)}.
\ee
We  emphasize that Jacobi elliptic functions are naturally functions of the variable $\beta$ on the {\it normalized} Jacobian $\mathbb J$ with quasi-periods $1, \tau = \frac {\varpi_1}{{\varpi_3}}$, while Weierstrass' $\wp, \zeta$ are doubly (quasi) periodic in the {\it unnormalized} variable $s$ and have periods $2{\varpi_3}, 2\varpi_1$: for this reason when mixing them in the same formula one should bear in mind that $s= 2{\varpi_3} \beta $ to translate from one to the other.

\subsection{Terminology.} 
The Jacobian of the elliptic curve \eqref{elliptic} is the torus $\mathbb J:= \C/\Z+\tau \Z$ and it provides the uniformization (i.e. the parametrization) of the algebraic curve \eqref{elliptic} via the Weierstrass' substitutions $w = \wp'(2{\varpi_3}  \beta), \ \ z = \wp(2{\varpi_3} \beta)$ as per \eqref{WP}. 
Points in the Jacobian $\mathbb J$ will be represented by points in the {\it fundamental domain}
\be
\label{fundamental}
\mathcal L:= \bigg\{
\Re \beta \in [0,1), \ \Im \beta \in \big[0, \Im (\tau) \big)\bigg\},
\ee
with $\beta=0$ corresponding to the point at infinity. 
We refer to $(w,z)$ in \eqref{elliptic} as the {\it Weierstrass' representation}. The three points $e_1,e_2,e_3$ correspond to the half periods $\frac \tau 2, \frac {\tau+1}2, \frac 1 2$ (respectively, modulo the lattice $\Lambda_\tau := \Z+\tau \Z$). This is a slightly non-standard correspondence due to the choice of $\scr A, \scr B$ cycles.

\begin{figure}
\begin{center}
\begin{tikzpicture}[scale=1.8]
\draw[fill=black!10!white] (0,0) to  (0,2.4)to  (4,2.4) to (4,0 ) to cycle;
\node [above] at (0,2.4) {$\tau$};
\node [above] at (4,2.4) {$1+\tau$};
\node [below] at (4,0) {$1$};
\draw [red, fill] (0.7,0) circle[radius=0.015] node [below]  { $\beta_j$}; 
\draw [dashed](0,1.2) to (4,1.2);
\draw [red, fill] (3.3,0) circle[radius=0.015] node [below]  { $\beta_j^\star$}; 
\draw [blue, fill] (1.5,1.2) circle[radius=0.015] node [below]  { $\beta_j$}; 
\draw [blue, fill] (2.5,1.2) circle[radius=0.015] node [below]  { $\beta_j^\star$}; 
\draw [dashed, red!40!black] (4/2,0) to (4/2,2.4);
\node at (4/2,-0.1) {$(e_3)$};
\draw [fill] (2,0) circle[radius=0.016];
\draw [fill] (0,1.2) circle[radius=0.016];
\draw [fill] (2,1.2) circle[radius=0.016];
\node at (2.2,1.31) {$(e_2)$};
\node at (-0.25,1.31) {$(e_1)$};
\end{tikzpicture}
\end{center}
\caption{The points $\beta_j$ in the Jacobian corresponding to the hot (red) solitons and cool (blue) solitons, and their corresponding involutions. Indicated in bracket the corresponding points on the elliptic curve.}
\label{Jacobian}
\end{figure}
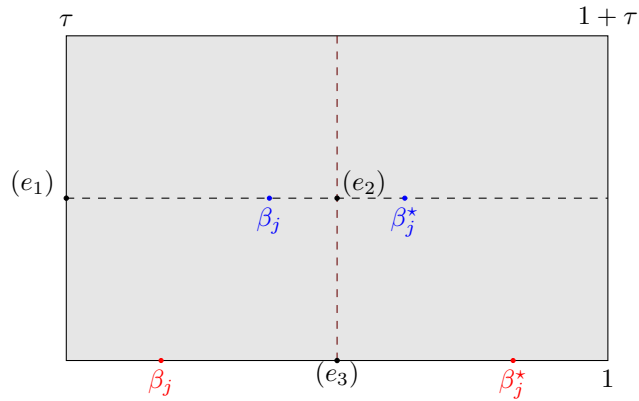

The points $b_j$ correspond to pairs of pre-images in the fundamental domain $\mathcal L$ of the Jacobian $\mathbb J$:
\be \label{beta.imp.def}
	b_j &= \wp( 2 {\varpi_3} \beta_j ) = \wp ( 2{\varpi_3} \beta^\star) ,  \qquad
	\beta_j \in \begin{cases} \le(0,\frac 1 2\ri) & 1\leq j \leq L \text{ (hot)} \\ \frac{\tau}{2}+ \le(0,\frac 1 2\ri) & L+1 \leq j \leq N \text{ (cool),} \end{cases}
\ee
where (see figure~\ref{Jacobian})
\begin{align}\label{beta.star.def}
\beta_j^\star = 1-\beta_j + \chi \tau, \qquad  \chi = \begin{cases} 0 & 1\leq j \leq L\\ 1 &L \leq  j \leq N. \end{cases}
\end{align}
The $\scr A_j, \scr B_j$ cycles on the curve \eqref{degRS} are defined in the following  way (see Fig. \ref{cycles}):
\begin{figure}
\begin{center}
\begin{tikzpicture}[scale=1.8]
\draw[fill=black!10!white] (0,0) to  (0,2.4)to  (4,2.4) to (4,0 ) to cycle;
\node [above] at (0,2.4) {$\tau$};
\node [above] at (4,2.4) {$1+\tau$};
\draw [line width=2pt, purple] (1,1.5) to ++ (40:0.2);
\draw [line width=2pt, blue] ($(1,1.5)+(135:0.04)$) to ++ (40:0.2);
\draw [blue, thick, postaction={decorate,decoration={markings,mark=at position 0.3 with {\arrow[line width=1.5pt]{>}}}}] (1.05, 1.58) circle  [radius=7pt];
\node [blue] at (1.3, 1.9)  {$\scr A_2$};
\draw [green!50!black, thick , postaction={decorate,decoration={markings,mark=at position 0.3 with {\arrow[line width=1.5pt]{>}}}}] (1.05, 1.58)  to [out=-45, in =135] node[pos=0.7, above ] {$\scr B_2$} (3.1, 0.99); 

\draw [line width=2pt, purple] (3,0.9) to ++ (40:0.2);
\draw [line width=2pt, blue] ($(3,0.9)-(135:0.04)$) to ++ (40:0.2);
\node [below] at (4,0) {$1$};
\draw [dashed](0,1.2) to (4,1.2);
\draw [dashed, red!40!black] (4/2,0) to (4/2,2.4);
\end{tikzpicture}
\end{center}
\caption{The curve \eqref{guide} before degeneration; the two small cuts are around the  images of the point $(Z=0, Y=\mp 4e_1e_2e_3)$, and their two sides  are identified as suggested by the color-coding.}
\label{degenerationJac}
\end{figure}
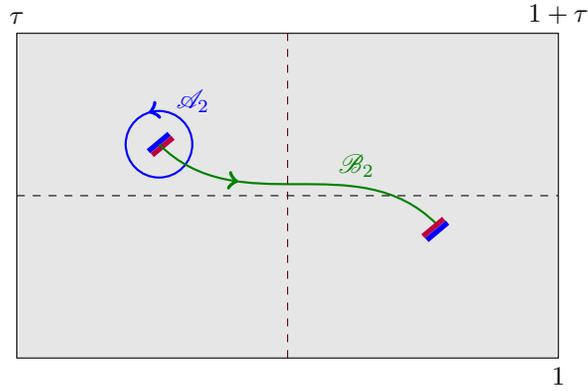

\begin{itemize}
\item for the bands around the hot solitons $b_j$ the $\scr A_j$ cycle is a small circle around the band and the $\scr B_j$ cycle is a contour from the right end of the small band to $e_3$ of the main band (on both sheets);
\item 
similarly, for the bands around the cool solitons $b_{L+j}$ the $\scr A_j$ cycle is a small circle around the band and the $\scr B_j$ cycle is a contour from the {\it left} end of the small band to $e_2$ of the main band;
\item  the cycles of the elliptic curve will be numbered $N+1$ and the $\scr A_{N+1}$ cycle and $\scr B_{N+1}$ as shown in Fig. \ref{cycles}. 
\end{itemize}

\subsection{General properties of nodal curves and their period matrix}
We refer to Chapter III of Fay for a general approach but here we give a perfunctory account of the ideas, tailored to the problem at hand, namely, that the limiting nodal curve is an elliptic curve. 

Consider, for guidance, the following example
\be
y^2 = (z^2-\varepsilon^2) 4(z-e_1)(z-e_2)(z-e_3), \ \ e_1 + e_2 + e_3=0. 
\label{guide}
\ee
As $\varepsilon\to 0$ the curve becomes a nodal elliptic curve; its {\it resolution} consists of an elliptic curve (represented in the canonical Weierstrass form)
\be
\le(\frac {y}z\ri)^2 = 4(z-e_1)(z-e_2)(z-e_3).
\ee
The functions $y, z$ on this elliptic curve can be parametrized as follows in terms of the Weierstrass' functions $\wp, \wp'$:
\be
z= \wp(s),\ \ y = \wp(s) \wp'(s)= \frac 12 \le(\wp(s)^2\ri)'.
\ee
This means that the algebra of  these functions separate all points of the elliptic curve {\it except} the two zeros of $\wp$; in other words the nodal curve is the result of identifying one pair of points in the limiting elliptic curve. 

The limiting elliptic curve without the identification of the two points is an example of {\it resolution} of a nodal curve; if we have several nodes the procedure is precisely the same:
\be
y^2 = \prod_{j=1}^N (z-\psi_j)^2 4(z-e_1)(z-e_2)(z-e_3)\ \ \Rightarrow\ \ \ 
y = \prod_{j=1}^{N} \le(\wp(s)-\psi_j\ri)^2 \wp'(s),\ \ \ z = \wp(s).
\ee
In the resolved elliptic curve we now have $N$ pairs of pairwise identified points; in particular the two points in each pair are interchanged by the elliptic involution (if they were in general positions, then this would not be the resolution of a hyperelliptic nodal degeneration).
Let us denote by $\beta_j, \beta_j^\star$ these pairs in the Jacobian, within the same fundamental domain \eqref{fundamental}. 

What remains of the corresponding $\scr A_j, \scr B_j$ cycles?\\
Even before degeneration we can represent the hyperelliptic curve as  the elliptic curve with small branch cuts around $\beta_j, \beta_j^\star$, pairwise identified, see for example Fig. \ref{degenerationJac}.
The $\scr A_j$ cycle is represented by  a small counterclockwise circle around one of the two pre-images of the node (say, $\beta_j$), while what survives of the $\scr B_j$ cycle is a  path joining the two pre-images of the given node in the resolved curve. 

If we have several nodes, of course the $\scr B_j$ cycles should be chosen as mutually non-intersecting paths joining the two pre-images of each node and staying within the same fundamental domain \eqref{fundamental}.

What happens then to the corresponding normalized holomorphic differentials, $\omega_k$, $k=1,\dots,N+1$, in the degeneration process?
As explained better in \cite{Fay}, they become the (unique) third kind differentials on the resolved curve with two simple poles at the pre-images $\beta_j, \beta_j^\star$ of the nodes $b_j$, residues $\frac 1{2i\pi} $ at $\beta_j$ and $-\frac 1{2i\pi}$ at $\beta_j^\star$, and vanishing $\scr A$-period on the resolved curve. They have the following form 
\be
\label{third}
\rho_{\beta_j, \beta^\star_j} := \frac 1 {2i\pi} \frac {\d}{\d \beta} \ln \frac {\theta_1(\beta-\beta_j)}{\theta_1(\beta-\beta_j^\star)} \d \beta.
\ee
The evident advantage of formulas such as \eqref{third} is that the differentials appear as total derivatives of (multivalued) functions on the curve and hence the integration is a pleasant experience. We  mention here that such formulas can be written for any genus Riemann surface (hyperelliptic or not) in terms of Riemann theta functions and in fact all the considerations here translate with no substantial difference to higher genus.

The matrix of $\scr B$ periods will have, in the off--diagonal entries, the following limit
\be
{\bs \Omega}_{\ell,k} =\oint_{\scr B_\ell}\omega_k \to \int_{\beta_\ell^\star}^{\beta_\ell} \rho_{\beta_k,\beta_k^\star} = \frac 1{2i\pi} \ln \frac {\theta_1(\beta_\ell^\star - \beta_k^\star)}
{\theta_1(\beta_\ell^\star - \beta_k)}
\frac 
{\theta_1(\beta_\ell - \beta_k)}
{\theta_1(\beta_\ell - \beta_k^\star)},
\label{Weyl}
\ee
where $\omega_k$ denotes the $k^\mathrm{th}$ normalized holomorphic differential, see figure~\ref{cycles}.
The above formula determines ${\bs \Omega}_{\ell, k}$ only up to integers; this phenomenon, rather than a nuisance, is a feature. It simply reflects the fact that we can add to a period matrix ${\bs \Omega}$ an arbitrary integer matrix by adding to the $\scr B$--cycles an integer combination of the $\scr A$--cycles.

We  also mention that the symmetry in the exchange $\ell,k$ is a manifestation of Weil reciprocity or as a consequence of Riemann bilinear relations.\\[3pt]
If now $\d \beta$ denotes the  normalized differential on the resolved elliptic curve, then in the limit we have   $\omega_{N+1}\to \d \beta$. 
It follows that the limit  of ${\bs \Omega}_{\ell, N+1}$  tends to  the difference of the Abel maps of the pairs $\beta_\ell, \beta_\ell^\star$:
\be
{\bs \Omega}_{\ell, N+1} =\oint_{\scr B_\ell} \omega_{N+1} \to \int_{\beta_\ell^\star}^{\beta_\ell}\d \beta = \beta_\ell-\beta_\ell^\star = \oint_{\scr B_{N+1}} \rho_{\beta_\ell,\beta_\ell^\star}.
\ee
In this case, the symmetry ${\bs \Omega}_{\ell, N+1} = {\bs \Omega}_{N+1,\ell}$ in the $\eps \to 0$ limit is simply a consequence of the Riemann bilinear identity on the elliptic curve.

We mention here that if the resolved curve were of higher genus, at this point we would simply obtain the full Abel map of the divisor of degree zero $\scr D_\ell= \beta_\ell -\beta_\ell^\star$, $\ell=1,\dots, N$.

\subsubsection{Properties of the period matrix.}
With the choice of $\scr A/\scr B$ cycles made above (see figure~\ref{cycles}), the matrix, ${\bs \Omega}$, of normalized $\scr B $--periods satisfies 
\begin{gather}
{\bs \Omega}_{\ell,m} = \oint_{\scr B_\ell} \omega_m ={\bs \Omega}_{m\ell},  \qquad \ell,m=1\dots N+1, \\
{\bs \Omega}_{ N+1, \ell}={\bs \Omega}_{\ell, N+1} \in \R, \qquad {\bs \Omega}_{N+1,N+1} \in i\R_+, \qquad  {\bs \Omega}_{\ell, m} \in i\R, \ \ell, m\leq N.
\end{gather}
We can summarize the discussion of the previous paragraph in the following Lemma.
\begin{lemma}
As $\varepsilon\to 0$ we have that 
\begin{align}
&\begin{aligned}
&\omega_j \to  \frac 1{2i\pi} \frac {\d}{\d \beta} \ln \frac {\theta_1(\beta - \beta_j)}{\theta_1(\beta+\beta_j -1)} \d \beta, &&j=1,\dots, L \quad \text{(hot solitons)}; \\
&\omega_{j} \to  \frac 1{2i\pi} \frac {\d}{\d \beta} \ln \frac {\theta_1(\beta - \beta_j)}{\theta_1(\beta+\beta_j -\tau-1)} \d \beta,  &&j=L+1,\dots, N \quad \text{(cool solitons)}; 
\end{aligned} \\
&\omega_{N+1} \to \d \beta.
\label{3rd}
\end{align}
The matrix of normalized $\scr B$--periods has the following limits:
\begin{enumerate}
\item
The ``hot-hot'' part ($j, \ell=1,\dots, L$) is given by 
\be
\label{hothot}
{\bs \Omega}_{j,\ell}=\oint_{\scr B_j} \omega_\ell  \to 
\frac 1{i \pi}  \ln\le| \frac{\theta_1(\beta_j-\beta_\ell)} {\theta_1(\beta_j+\beta_\ell)} \ri| 
\qquad j \neq \ell
\ee
\item
The ``cool-cool'' part ($j, \ell=L+1,\dots, N$) is given by 
\be
\label{coolcool}
{\bs \Omega}_{j,\ell}=\oint_{\scr B_j} \omega_\ell  \to 
 &=
\frac 1{i \pi}  \ln\le| \frac{\theta_1(\beta_j-\beta_\ell)} {\theta_1(\beta_j+\beta_\ell -\tau)} \ri|
\qquad j \neq \ell
\ee
\item The ``hot-cool'' ($j=1,\dots, L$, $\ell = L+1,\dots, N$)  (or ''cool-hot'') part is given by 
\be
\label{hotcool}
{\bs \Omega}_{j,\ell}=\oint_{\scr B_j} \omega_\ell  \to 
 &=
\frac 1{i \pi}  \ln\le| \frac{\theta_1(\beta_j -\beta_\ell)} {\theta_1(\beta_j+\beta_\ell -\tau)} \ri|,\ \ \  j\in \{1,\dots, g\}, \ \ \ell \in \{g+1,\dots, N\}
\ee
\item The hot and cool solitons interact with the finite genus background by 
\be
\label{hotabel}
{\bs \Omega}_{\ell,N+1} &\to  2\beta_\ell-1\in \R,  &\ell = 1,\dots, g
\\
\label{coolabel}
{\bs \Omega}_{\ell,N+1} &\to  2\beta_\ell-\tau-1\in \R , &\ell = g+1,\dots, N.
\ee
\item
Finally,
\be
\label{diagonal}
{\bs \Omega}_{jj} = i \ln \frac 1 \varepsilon + \mathcal O(1) \to +i\infty, \ \ \ j=1,\dots, N;  \ \ \ {\bs \Omega}_{N+1,N+1} \to \tau.
\ee

\end{enumerate}
\label{asympO}
\end{lemma}
\noindent 

\paragraph{The Riemann Theta function and its degeneration.}
Let $\Theta$ be the Riemann Theta function \cite{Fay};
\def\n{{\bf n}}
\be
\Theta\le(\bs X;  {\bs \Omega}\ri):= \sum_{{\bs \nu} \in \Z^{N+1} } {\rm e}^{i\pi {\bs \nu}^\transpose {\bs \Omega} {\bs \nu} + 2i\pi {\bs \nu}^\transpose {\bs X}}, \ \ \ {\bs X}\in \C^{N+1}
\ee
According to  Lemma \ref{asympO} we  partition the matrix ${\bs \Omega}$ into blocks:
\begin{gather}\label{Omega.blocks}
{\bs \Omega} = \le[
\begin{array}{c|c}
\mathbb B & {\bs \mu}\\
\hline
{\bs\mu}^\transpose & \Omega_{N+1,N+1}
\end{array}
\ri]
\end{gather}
where the diagonal of $\mathbb B$ diverges to $+i\infty$ (at the rate of $\ln \varepsilon$, but this is not important here). Recalling \eqref{beta.star.def}
 the limit of the matrix $\mathbb B_{\ell, j}$ can be uniformly written as 
\begin{gather}
\mathbb B_{\ell, j} \to \frac 1{i\pi} \ln \le|
\frac{\theta_1 (\beta_j - \beta_\ell)}{\theta_1(\beta_j - \beta_\ell^\star)}
\ri|, \qquad \mu_\ell \to \beta_\ell - \beta_\ell^\star = \le\{
\begin{array}{cc}
2\beta_\ell-1& \ell \leq L \text { (hot solitons)}\\
2\beta_\ell - \tau -1 & \ell \geq L+1 \text { (cool solitons)}.
\end{array} \ri.
\label{Blm}\\
\Omega_{N+1, N+1}\to \tau\nn
\end{gather}
Here the different form of the involution for $\beta_\ell$'s is due to the choice of fundamental domain \eqref{fundamental}.
With these preparations we can state and prove the main theorem:
\bt
\label{main}
Let us denote $\bs u = (1,1,\dots, 1,0)\in \C^{N+1}$,  ${\bs X} = [\bs{\psi},\beta] \in \C^{N+1}$. Then 
\be
\lim_{\eps\to 0_+} \Theta\le( {\bs X} - \frac 1 2 {\bs \Omega}(\eps) {\bs u} ; {\bs \Omega}(\eps) \ri)  = \det\le[
\1_N + \mathbb G\ri]
\theta_3\le(\beta - \mathcal A\ri),
\\
\mathbb G= [\mathbb G_{\ell m}]= \le[ 
\frac {
\ds 
\theta_3\le(\beta_\ell - \beta^\star_m + {\beta} -  \mathcal A\ri)}{ 
\ds
\theta_1\le( \beta_\ell - \beta^\star_m \ri) \theta_3\le(\beta -  \mathcal A\ri) }
\sqrt{C_\ell C_m} {\rm e}^{i\pi(\psi_\ell+\psi_m)}
\ri]_{\ell, m=1}^N \label{Noice}
\ee
where $\beta_j, \beta_j^\star$, defined in \eqref{beta.imp.def}-\eqref{beta.star.def}, are the {two pre-images of $b_j$ in the Jacobian of the resolved elliptic curve},  and 
\be
\mathcal A=\frac 1 2 \sum_j \le(\beta_j-\beta^\star_j\ri),\qquad
C_\ell:=  \theta_1(\beta_\ell - \beta_\ell^\star  )
\prod_{\substack{ k\in \{1,\dots, N\} \\ k \neq \ell }}
{ \le| \frac 
  {\theta_1(\beta_k-\beta^\star_\ell)}
  {\theta_1(\beta_k-\beta_\ell) }
\ri|}.
\ee
\et

\begin{figure}
\includegraphics[width=0.99\textwidth]{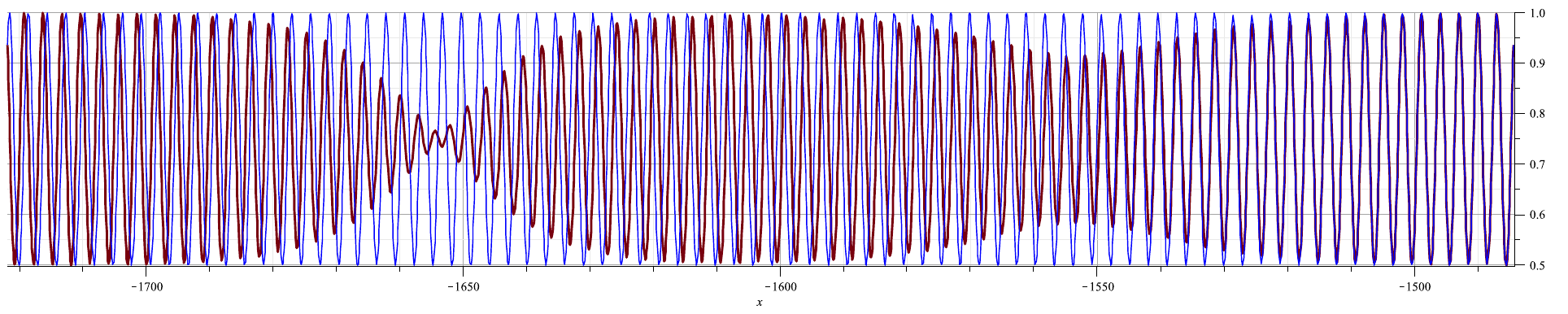}
\includegraphics[width=0.99\textwidth]{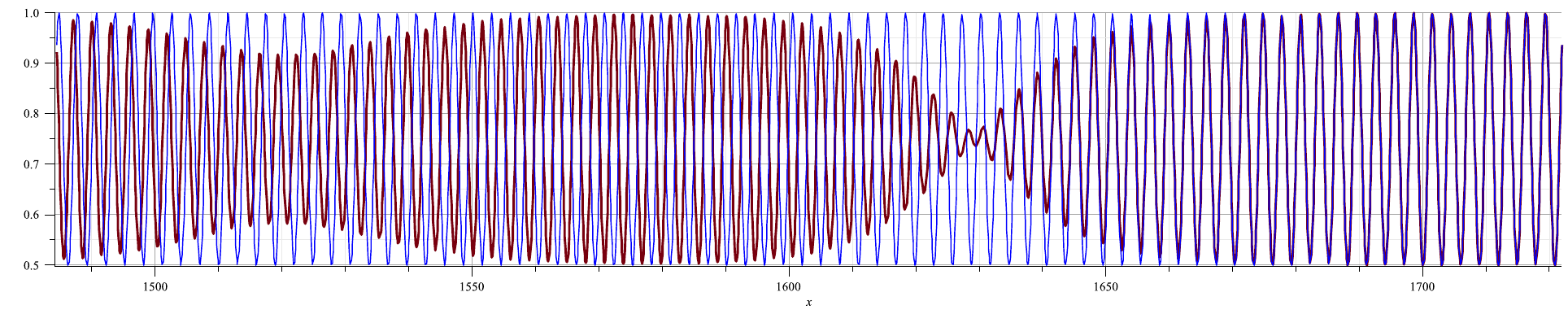}
\caption{A plot overlaying the 2-soliton solution (red: in this case two retrograde solitons) with the background cnoidal wave. The observable phase shift of the background that occurs from the right to the left of each solitary disturbance, is described in Theorem \ref{scatteringthm}. }
\label{2dimshift}
\end{figure}

\begin{figure}
\includegraphics[width=0.99\textwidth]{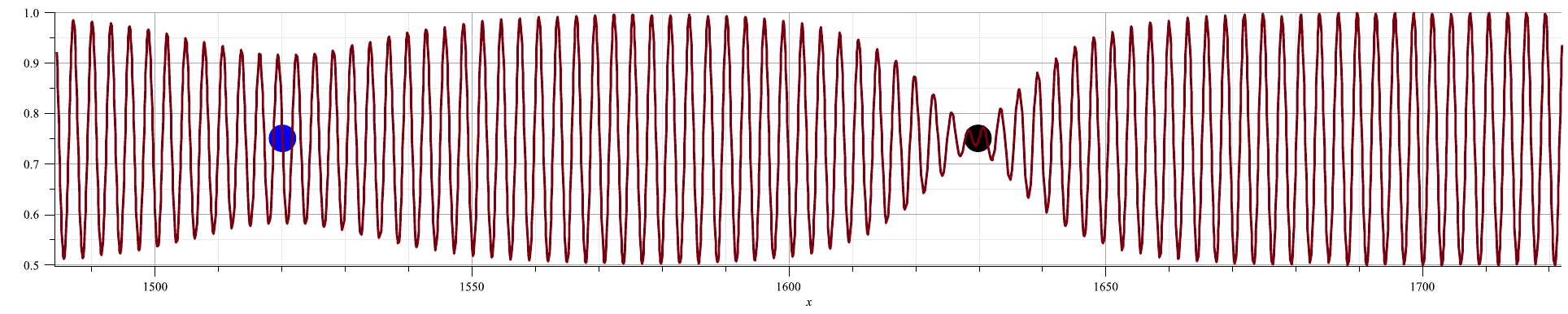}
\includegraphics[width=0.99\textwidth]{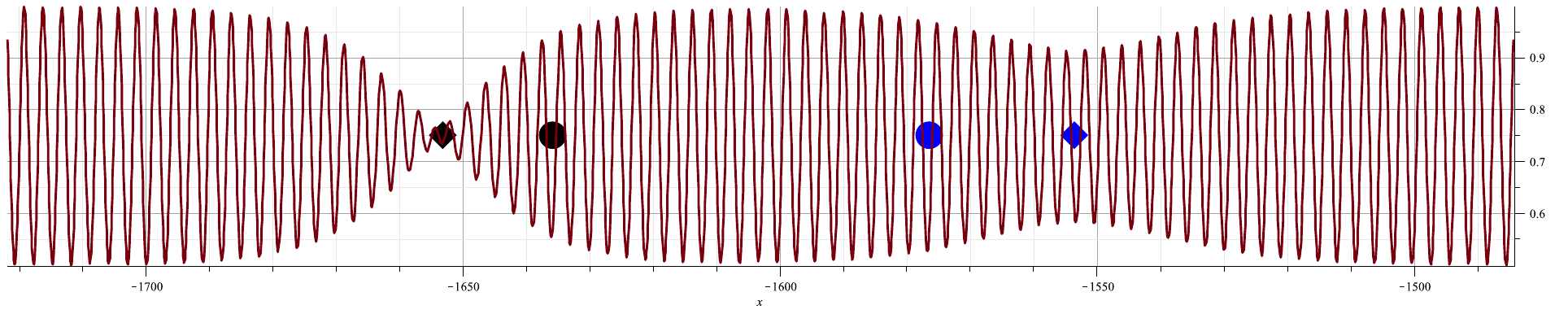}
\caption{The same setup as in Fig. \ref{2dimshift}. The top pane shows two retrograde solitons for a large negative time (remote past); the dots indicate, for reference, the ``position'' of the two disturbances. 
Both disturbances have negative velocity (move to the left); the narrow{-waist}, disturbance travels with greater speed than the wider{-waist} disturbance. 
In the remote future (below) the narrower disturbance has overtaken the wider one.  The circles indicate where the two disturbances would be if propagating on the pure cnoidal background; the position of the diamonds is computed by the shift given by the formula in Theorem \ref{scatteringthm}. 
The numerical values for this example are as follows: $e_1=2, e_2=1, e_3=-3$, $\beta_1= 1/4+\frac \tau 2, \ \ \beta_2 = 0.36 + \frac \tau 2$, $V_1\simeq -8.94427, \ \ V_2 \simeq-8.4810443$, $\Delta(\beta_1,\beta_2) \simeq -17.32$,   $\Delta(\beta_2,\beta_1) \simeq 22.878$, and the times shown above are $t_\pm \simeq  \pm 182.5586$.}
\label{2dimshiftreference}
\end{figure}

\noindent {\bf Proof of Theorem \ref{main}.}
The proof follows similar steps to the proof for ordinary solitons by Mumford in \cite{Mumford}, where, however, the author ends up with a different determinantal formula (not a Fredholm determinant). 
Here we also have  the added twist that at some point we need a special identity of determinants due to Fay \cite{Fay}.
With the established notations and splitting the summation integer vector ${\bs \nu} = [{\bs n}, m] \in \Z^{N+1}$,  $\bs 1 = (1,1,1,\dots, 1)\in \R^N$ we have 
\bea
\Theta\le( {\bs X} - \frac 1 2 {\bs \Omega} {\bs u} \ri) 
&=
\sum_{m\in \Z} \sum_{\n\in \Z^N}
\exp i\pi \le[
   m^2 \Omega_{N+1,N+1} +  \n^{\transpose} \mathbb B \n
+2 m {\bs \mu} ^{\transpose} \n + 
2\le( 
m \beta 
+ \n^{\transpose}\bs{\psi} 
 - 
 \frac{1}{2} \n^{\transpose} \mathbb B {\bs 1}
 -\frac m2 \bs \mu ^{\transpose} \bs 1\ri)
\ri]\nn
=\\
= \sum_{\n\in \Z^N}\sum_{m\in \Z} &
\exp i\pi\le[ m^2 \Omega_{N+1,N+1}+\sum_\ell (n_\ell^2-n_\ell)\mathbb B_{\ell \ell}
+ \sideset{}{'}\sum_{\ell ,k} n_\ell (n_k-1) \mathbb B_{\ell k} 
+ 2m\le(\bs \mu^{\transpose} \n + \beta - \frac 12 \bs \mu^{\transpose} \bs 1\ri)+ 2 \n^{\transpose} \bs{\psi}
\ri].\nn
\eea
Since $\mathbb B_{\ell,\ell}\to + i\infty$, in the limit only the vectors $\n$ with entries $n_\ell \in \{0,1\}$ will contribute to the sum, while all the other being suppressed (we leave the details of the use of dominated convergence to the reader).
Thus the limit of the above sum as $\varepsilon \to 0$ yields (the prime on the summation indicating that the sum is for $\ell \neq k$)
\begin{multline}
\lim_{\varepsilon \to 0} \Theta\le( {\bs X} - \frac 1 2 {\bs \Omega} {\bs u} \ri) 
= \\
\lim_{\varepsilon \to 0} 
\sum_{\n\in \{0,1\}^N}\sum_{m\in \Z}
\exp i\pi\le[  m^2 \Omega_{N+1,N+1} +\sideset{}{'}\sum_{\ell ,k} n_\ell (n_k-1) \mathbb B_{\ell k} 
+ 2m\le(\bs \mu^{\transpose} \n + \beta - \frac 12 \bs \mu^{\transpose} \bs 1\ri)+ 2 \n^{\transpose} \bs{\psi} 
\ri]=
\\
=
\lim_{\varepsilon \to 0} 
\sum_{\n\in \{0,1\}^N}
\exp i\pi\le[\sideset{}{'}\sum_{\ell ,k} n_\ell (n_k-1) \mathbb B_{\ell k} + 2 \n^{\transpose} \bs{\psi}
\ri]\theta_3\le( \bs \mu^{\transpose} \n + \beta - \frac 12 \bs \mu^{\transpose} \bs 1\ri) 
\end{multline}
We now use the results \eqref{Weyl}
\be
&
\nn
\lim_{\varepsilon \to 0} \Theta\le( {\bs X} - \frac 1 2 {\bs \Omega} {\bs u} \ri) 
=\\
&=\sum_{\n\in \{0,1\}^N}
\sideset{}{'}\prod_{\ell,k}
\le(
 \frac {\theta_1(\beta_k-\beta_\ell)\theta_1(\beta_k^\star-\beta^\star_\ell)}
 {\theta_1(\beta_k-\beta^\star_\ell)\theta_1(\beta^\star_k-\beta_\ell)}
 \ri)^{\frac{n_\ell (n_k-1)}2}
 {\rm e}^{2i\pi \sum_\ell n_\ell \psi_\ell}
\theta_3
\le( \beta +\sum_\ell \le(\beta_\ell-\beta_\ell^\star\ri) (n_\ell - 1/2)  
\ri).
\label{139}
\ee
Note that the expression $\frac {\theta_1(\beta_k-\beta_\ell)\theta_1(\beta_k^\star-\beta^\star_\ell)}
 {\theta_1(\beta_k-\beta^\star_\ell)\theta_1(\beta^\star_k-\beta_\ell)}=\lim_{\varepsilon\to 0} {\rm e}^{i\pi {\bs \Omega}_{\ell k}}$ is a {\it positive} real number.
Now we equivalently sum over all possible subsets $S\subset\{1,2,\dots, g\}$ consisting of increasing indices.
\begin{equation}
\begin{aligned}
\lim_{\varepsilon \to 0} \Theta\le( {\bs X} - \frac 1 2 {\bs \Omega} {\bs u} \ri) 
&=
\sum_{\substack{ S\subset\{1,\dots ,N\} \\ S\text{ ordered}}}
\sideset{}{'}\prod_{\ell<k \in S}
\frac {\theta_1(\beta_k-\beta_\ell)\theta_1(\beta_k^\star-\beta^\star_\ell)}
 {\theta_1(\beta_k-\beta^\star_\ell)\theta_1(\beta^\star_k-\beta_\ell)}
\times
\\
&\times
\prod_{\ell \in S}
\le(
{\rm e}^{2i\pi \psi_\ell}
\prod_{\substack{
 k\in \{1,\dots, N\} \\ k \neq \ell}} \le|  
 \frac 
 {\theta_1(\beta_k-\beta^\star_\ell)\theta_1(\beta^\star_k-\beta_\ell)}
 {\theta_1(\beta_k-\beta_\ell)\theta_1(\beta_k^\star-\beta^\star_\ell)}
 \ri|^\frac 1 2\ri)
\theta_3\le( \beta +\sum_{\ell\in S} \le(\beta_\ell-\beta_\ell^\star\ri)-\frac 1 2 \sum_{j} \le(\beta_j-\beta_j^\star\ri)  \ri)
\label{130}
\end{aligned}
\end{equation}
We therefore conclude that 
\be
\lim_{\varepsilon \to 0} \Theta\le( {\bs X} - \frac 1 2 {\bs \Omega} {\bs u} \ri)  =&
\sum_{\substack{S\subset\{1,\dots ,N\} \\S\text{ ordered}}}
\sideset{}{'}\prod_{\ell<k \in S}
\frac {\theta_1(\beta_k-\beta_\ell)\theta_1(\beta_k^\star-\beta^\star_\ell)}
 {\theta_1(\beta_k-\beta^\star_\ell)\theta_1(\beta^\star_k-\beta_\ell)}
\prod_{\ell \in S}
\frac{C_\ell \, {\rm e}^{2i\pi \psi_\ell}}{\theta_1(\beta_\ell-\beta_\ell^\star)} \,
\theta_3\le( \beta +\sum_{\ell\in S}(\beta_\ell-\beta_\ell^\star) - \mathcal A \ri)\ ,
\label{131}
\ee
where we have set for brevity (the {\it norming constants}):
\be
C_\ell := \theta_1(\beta_\ell - \beta_\ell^\star  )
\prod_{\substack{ k\in \{1,\dots, N\} \\ k \neq \ell }}
 \le|\frac
 {\theta_1(\beta_k-\beta^\star_\ell)\theta_1(\beta^\star_k-\beta_\ell)}
  {\theta_1(\beta_k-\beta_\ell)\theta_1(\beta_k^\star-\beta^\star_\ell)}
  \ri|^{\frac 1 2}
 ,\qquad
\mathcal A =\frac 1 2  \sum (\beta_j-\beta_j^\star).
\ee
We now need the Fay identity \cite[p.~33]{Fay}. In genus one it can be formulated as the  identity \eqref{Fay} below, where $n\in \N$ is  arbitrary; $x_1,\dots, x_n, \hat x_1,\dots, \hat x_n$ are arbitrary points on the elliptic curve, and $\mathcal E\in \C$ is any value for which $\theta_3 (\mathcal E)\neq 0$:
\be
\label{Fay}
\det \le[\frac {\theta_3\le(x_\ell - \hat{x}_m + \mathcal E \ri)}{\theta_1\le(x_\ell - \hat{x}_m\ri)
\theta_3\le( \mathcal E \ri)}\ri]_{\ell, m=1}^n
=
\frac{\theta_3\le(\sum_j (x_j-\hat{x}_j) + \mathcal E\ri)}{\theta_3\le( \mathcal E \ri)}
\frac {\ds
\prod_{j<k} \theta_1(x_j-x_k)\theta_1(\hat{x}_k-\hat{x}_j)
}{\ds \prod_{k,j=1}^n \theta_1(x_j-\hat{x}_k)}.
\ee
If we specialize to $x_j = \beta_j$, $\hat{x}_j = \beta_j^\star$,  $j=1,\dots, N$ and $\mathcal E = \beta - \mathcal A$  and let $\sharp(S)$ denote the cardinality of the set $S$, then we obtain 
\be
\label{Fayb}
\frac{\ds \theta_3\le(\sum_{\ell\in S}(\beta_\ell-\beta_\ell^\star) + \beta - \mathcal A\ri)}{\theta_3\le( \beta - \mathcal A \ri)}
\frac {\ds
\prod_{\ell<k \in S} \theta_1(\beta_\ell-\beta_k) \theta_1 (\beta^\star_k-\beta^\star_\ell)
}{\ds \prod_{\ell, k \in S}^{\sharp(S)}  \theta_1(\beta_\ell-\beta^\star_k)}
=&
\det \le[\frac {\theta_3\le(\beta_\ell - \beta^\star_m +  \beta - \mathcal A\ri)}{\theta_1\le(\beta_\ell -\beta^\star_m\ri)
\theta_3\le( \beta- \mathcal A \ri)}\ri]_{\ell, m\in S}.
\ee
The rhs of \eqref{Fayb} reads
\be
({\rm rhs} \ \ref{Fayb})=&
\frac{\theta_3\le(\ds \beta +\sum_{\ell\in S}(\beta_\ell-\beta_\ell^\star) - \mathcal A\ri)}{\theta_3\le( \beta - \mathcal A \ri)}
\prod_{\ell\in S}
\frac 1{\theta_1(\beta_\ell-\beta_\ell^\star)}
\prod_{\ell<k \in S}
\frac{ \theta_1(\beta_\ell-\beta_k) \theta_1(\beta^\star_k-\beta^\star_\ell)
}
{{\ds  \theta_1(\beta_\ell-\beta^\star_k)\theta_1(\beta_k-\beta^\star_\ell)}}.
\ee
Thus we find that \eqref{131} becomes
\be
\lim_{\varepsilon \to 0} \Theta\le( {\bs X} - \frac 1 2 {\bs \Omega} {\bs u} \ri) 
&= \theta_3\le( \beta  - \mathcal A\ri) \sum_{\substack{S\subset\{1,\dots ,N\} \\ S\text{ ordered}}}
\det \le[\frac {\theta_3\le(\beta_\ell - \beta^\star_m + \beta - \mathcal A\ri)}{\theta_1\le(\beta_\ell -\beta^\star_m\ri)
\theta_3\le(v- \mathcal A \ri)}
 \sqrt{C_\ell C_m}{\rm e}^{i\pi( \psi_\ell + \psi_m)}\ri]_{\ell, m\in S}
=\\
&=\theta_3\le( \beta  - \mathcal A\ri) \sum_{\substack{S\subset\{1,\dots ,N\} \\ S\text{ unordered}}} 
\frac 1{{\sharp}(S)!}
\det \le[\frac {\theta_3\le(\beta_\ell - \beta^\star_m + \beta - \mathcal A\ri)}{\theta_1\le(\beta_\ell -\beta^\star_m\ri)
\theta_3\le( \beta - \mathcal A \ri)} 
 \sqrt{C_\ell C_m}{\rm e}^{i\pi( \psi_\ell + \psi_m)}\ri]_{\ell, m\in S}.
\ee
This last summation is precisely the Fredholm expansion of the determinant \cite{SimonIdeals} in the statement.
\QED
\subsection{KdV tau function}
In the finite-gap integration method the vector $\bs X$ (the argument of Riemann's theta function) evolves linearly with respect to $x, t$:
\be
\bs X = x \frac{\bs P }{2\pi}+ t \frac{\bs E}{2\pi} + \bs X_0.
\ee
With our choices of the $\scr A, \scr B$ cycles the first $N$ components are purely imaginary and the last component is real. The vector ${\bs X_0}$ encodes the initial conditions and it is arbitrary (with the same reality properties).
The vectors $\bs P, \bs E$ are the vectors of $\scr B$--periods of second--kind differentials  on the Riemann surface \eqref{degRS}:
\be \label{P/E vectors}
P_\ell = \oint_{\scr B_\ell} \d P;\ \ \ E_\ell = \oint_{\scr B_\ell} \d E
\ee
where the {\it quasi-momentum} and {\it quasi-energy} differentials are the unique differentials of the second kind (i.e. without residues) with a single pole at $\infty$ of order $2$ and $3$, respectively, normalized to have vanishing $\scr A$--periods on the hyperelliptic curve, with the following prescribed singular behaviour:
\be
\d P(z) = \le(\frac 1 {4\sqrt z} + \mathcal O\le(z^{-\frac 32} \ri) \ri)\d z;\ \ \  \oint_{\scr A_j}\d P=0, \ \ j=1\dots N+1
\\
\d E(z) =  \le(\frac {3} 4  \sqrt z + \mathcal O\le(z^{-\frac 3 2}\ri)\ri)\d z
\ \ \ 
\oint_{\scr A_j} \d E=0, \ \ j=1,\dots, N+1.
\ee
Here $\sqrt z$ means the root with the branchcut along $\R_+$ and $\sqrt{z_+}\in \R_-$.
In particular, we have that the Abelian integrals behave as 
\be
x\int\d P + t \int \d E \simeq x\frac{ \sqrt{z}}2 + t \frac {z^\frac 3 2}2.
\ee
The complete formula for the tau function of the KdV solution contains also \cite{Dubrovin} an exponential as follows
\be
\label{taugen}
{\bs \tau}(x,t) = {\rm e}^{-x^2 C} \Theta\le( \frac{x{\bf P}+ t{\bf E}}{2\pi} + {\bs X_0}\ri),
\ee
and the finite-gap solution of the KdV equation is given by 
\be
u(x,t) = 2\frac {\pa^2}{\pa x^2} \ln {\bs \tau}(x,t),
\ee
which satisfies the KdV equation with the following coefficients:
\be
u_t  + u_{xxx} + 6uu_x=0 . 
\label{KdV} 
\ee
In fact, the exponential in \eqref{taugen} contains a quadratic form in all times of the hierarchy, but this has no effect on the solution $u(x,t)$ of KdV, due to the logarithmic differentiation in $x$.
However, the constant $C$ is essential. Expressions for it  can be found in \cite{FFM} but we need a different description here \cite{Dubrovin, BertoNotes} which involves the canonical bi-differential of the Riemann surface. 

On any Riemann surface of genus $g$ the canonical bi-differential ${\bf B}(p,q)$  is a differential in both arguments that satisfies 
\begin{enumerate}
\item ${\bf B}(p,q)= {\bf  B}(q,p)$;
\item As a differential in $p$, $\mathbf{B}(p,q)$ has a unique double pole for $p=q$ (and viceversa), with bi-residue $1$ in the sense that for any local coordinate $\xi$  we have
\be
{\bf B}(p,q) = \frac { \d w \d s}{(w-s)^2} (1 + \mathcal O(w-s)^2), \ \ w = \xi(p),  \ s = \xi(q).
\ee
\item Integration with respect to $p$ along the $\scr A$--cycles yields zero, identically in $q$ (and viceversa).
\end{enumerate}
Formulas for this object in terms of Theta functions can be found in \cite{Fay}, Ch. II, but  here we do not need  any further detailed information.

The constant $C$ appearing in \eqref{taugen} is then given by \cite{Dubrovin}\footnote{The theorem in loc. cit. is  Theorem 3.6.15 on page 138, stated without proof (and for the KP tau function), which is given as a collection of exercises. The solution to these exercises is contained then in \cite{BertoNotes}.}
\be
C = \frac 1 2 \res{p=\infty}\res{q=\infty} \frac {\sqrt{z(q)} }2\frac{\sqrt{z(p)}}2 {\bf B}(p,q)  =  \frac 1 2 \res{q=\infty} \frac {\sqrt{z(q)} }2 \d P(q)
\ee
where  the Riemann surface is the hyperelliptic surface of the form \eqref{degRS}.

The only information that we need here is that under the degeneration of the curve the bi-differential reduces to the corresponding bi-differential on the elliptic resolution of the limiting curve \cite{Fay}, Ch. IV.

Thus we need to figure out what is ${\bf B}(s,w)$ for an elliptic curve. 
A simple verification shows
\be
{\bf B}(s,w) = \le(\wp\le(s-w\ri) + \frac {\zeta({\varpi_3})}{{\varpi_3}}\ri) \d w \,\d s
\ee
Since $z = \wp(s) = \frac 1 {s^2} + \mathcal O(1)$, we see that the above residue becomes
\be
C = \frac 1 {{8}} \res{s=0}\res{w=0}\frac 1 {sw}\le(\wp\le(s-w\ri) + \frac {\zeta({\varpi_3})}{{\varpi_3}}\ri) \d w \d s = \frac {\zeta({\varpi_3})}{{8}{\varpi_3}}
\ee
\subsubsection{KdV evolution in  the degeneration}

We can identify the differentials of the quasi-momentum and quasi-energy in the limit   as $\varepsilon\to 0$;  in fact if they are normalized to have vanishing $\scr A$--cycles (where the $\scr A$ cycles are the ones around the nodal degenerations),  in the limit they  tend to the corresponding differentials on the elliptic curve (the resolution of the nodal hyperelliptic curve).  
\bl
The quasi-momentum $\d P$ and quasi-energy $\d E$ on the elliptic curve are 
\be
\label{Qmom}
\d P(s) &=-{\frac 1 2} \le(\wp(s) +  \frac{\zeta\le({\varpi_3} \ri)}{{\varpi_3}}\ri)\d s ={\frac 1 2}\frac 1{ (2{\varpi_3})^2 }\frac {\d^2}{\d s^2} \ln \theta_1
\le(\frac {s}{2{\varpi_3}};\tau\ri)   \ \ \ \Rightarrow\ \ \ 
\\&
\label{2.67}
P(s)  = \int_{\varpi_3}^s \d P= {\frac 1 2}   \le(\zeta(s) - s  \frac{\zeta\le({\varpi_3}\ri)}{{\varpi_3}}\ri)  = \frac 1{{4}{\varpi_3}} \frac{\theta_1'\le(\frac {s}{2{\varpi_3}};\tau\ri)}{\theta_1\le(\frac {s}{2{\varpi_3}};\tau\ri)}
\\
\label{Qen}
\d E(s) &= -{\frac 1 4} \wp''(s)\d s\ \ \ \ \Rightarrow \ \ \ E(s) \ =\int_{\varpi_3}^s \d E = -{\frac 1 4}  \wp' (s).
\ee
\el
\noindent {\bf Proof.}
We can express them in terms of Weierstrass elliptic functions. Since $z = \wp(s) = \frac 1 {s^2} + \mathcal O(1)$, 
\be
\d P =\le(\frac 1{4\sqrt{z}} + \mathcal O\le(z^{-\frac 32}\ri)\ri)\d z= -\le( \frac 1{2s^2}  + \mathcal O(1) \ri)\d s 
\ee
It follows that 
$$
\d P = - {\frac 1 2} (\wp(s) -C)\d s
$$
 with the constant chosen so as to  have vanishing $\scr A$-period:
\be
2{\varpi_3} C = \int_{\varpi_1}^{\varpi_1+2{\varpi_3} } \wp(s) \d s = \zeta\le(\varpi_1\ri) - \zeta\le(\varpi_1+2{\varpi_3}\ri) = - 2\zeta\le({\varpi_3}\ri)\ \ \ \ \ \in \R
\ee
where we have used \eqref{periodzeta}.
Thus the quasi-momentum is given by \eqref{Qmom}. 
Similarly 
\be
\d E = {\frac 1 2} \le({-3}  \frac 1{s^4} + \mathcal O(1) \ri)\d s.
\ee
Formula \eqref{Qen} then  follows.
\QED

\bl\label{lem-PE}
The quasi-momenta and quasi-energies of the solitons are 
\be
P_j & = \le\{
\begin{array}{cc}
\ds \zeta(2{\varpi_3}\beta_j) - 2 \zeta\le({\varpi_3}\ri)\beta_j \in i\R_+& j \leq L \\[10pt]
\ds   
\zeta(2{\varpi_3}\beta_j)  -2\beta_j \zeta\le({\varpi_3}\ri) +\frac{ \pi i}{2{\varpi_3}}
\in i\R_+& j \geq L+1\\
\end{array}
\ri. 
\label{Pjs}
\\
\label{Ejs}
E_j & = \le\{
\begin{array}{cc}
\ds=-\frac 12 \wp'({2{\varpi_3}}\beta_j) = - \sqrt{(b_j-e_1)(b_j-e_2)(b_j-e_3)} \in i\R_- & j \leq L\\[10pt]
\ds=-\frac 1 2 \wp'({2{\varpi_3}}\beta_j)= -\sqrt{(c_j-e_1)(c_j-e_2)(c_j-e_3)} \in i\R_+ & j \geq L+1\\
\end{array}
\ri. 
\\
\label{EP-last}
&E_{N+1}=0, \ \ P_{N+1} = -\frac {i\pi}{{2}{\varpi_3}} \in \R_+
\ee
Alternatively we can write the quantities $P_j$ in  uniform way using \eqref{Qmom} 
\be
\label{Pjalt}
P_j = P(2{\varpi_3} \beta_j)- P(2{\varpi_3}\beta_j^\star) =
\frac 1{{4}{\varpi_3}} \le(
\frac{\theta_1'\le(\beta_j\ri)}{\theta_1\le(\beta_j\ri)} 
-
\frac{\theta_1'\le(\beta^\star_j\ri)}{\theta_1\le(\beta^\star_j\ri)} 
\ri)
=\frac 1{{2} {\varpi_3}} 
\frac{\theta_1'\le(\beta_j\ri)}{\theta_1\le(\beta_j\ri)}  +\chi\frac {i\pi}{{2}{\varpi_3}}
\ee
where $\chi=1$ for the cool solitons and $\chi=0$ for the hot ones.
\el
\noindent{\bf Proof.}
We use \eqref{Qmom} and start from $N+1$:
\be
P_{N+1} = P\le({\varpi_3} + 2\varpi_1\ri)- P\le({\varpi_3}\ri) =  
  \zeta \le(\varpi_1\ri)- \frac{ \varpi_1}{{\varpi_3}}\zeta \le({\varpi_3}\ri) = \frac {-i\pi}{{2}{\varpi_3}}
 \ee
where we have used one of Legendre's relations \cite[\href{http://dlmf.nist.gov/23.2.E14}{Eq. 23.2.14}]{DLMF}.
For $j\leq L$ (hot solitons) we have 
\begin{multline}
P_j = P(2{\varpi_3}\beta_j) - P(2{\varpi_3}\beta_j^\star) 
=
{\frac 1 2}\zeta(2{\varpi_3}\beta_j) - {\frac 1 2}\zeta(2{\varpi_3}(1-\beta_j)) - {2}  \beta_j \zeta\le({\varpi_3}\ri) +  \zeta\le({\varpi_3}\ri)
\\
=\zeta(2\varpi_3\beta_j)  - 2 \beta_j \zeta\le({\varpi_3}\ri) 
\end{multline}
where we have used the periodicity of $\zeta$ \eqref{periodzeta} together with its oddness.
For the cool  solitons we have 
\be
P_j = P({2{\varpi_3}} \beta_j) - P({2{\varpi_3}}(\tau+1-\beta_j)) =
\zeta(2{\varpi_3}\beta_j)  -2\beta_j \zeta\le({\varpi_3}\ri) +\frac{ \pi i}{{2}{\varpi_3}}
\ee
The formula \eqref{Ejs} is similarly obtained evaluating $E(\beta_j) - E(\beta_j^\star) = 2E(\beta_j)$: note that $\d E$  is an exact differential; this reflects the well known fact that the $1$--gap solution is stationary (up to Galilean invariance).
\QED

We summarize the computation in the following theorem.
\bt
\label{solitons!}
In the limit $\varepsilon \to 0$ the vector of phases 
has the following limit:
\begin{equation}
 {\bs X}(x,t)\to { \begin{bmatrix}  \psi_1(x,t) \\ \vdots \\ \psi_N(x,t) \\ \beta(x,t) \end{bmatrix} }
\qquad
\begin{aligned}\label{XwithPhases}
\psi_j(x,t) &:= {(x-x_j^{(0)})}\frac{P_j}{2\pi} + t \frac {E_j}{2\pi} , \quad j=1, \dots, N\\
\ds \beta(x)&:= \frac{{x-x_0}}{{4}i{\varpi_3}} 
\\ P_j &\in i\R_+, \quad E_{1,\dots, L}\in i\R_-, \quad  E_{L+1,..., N}\in i\R_-.
\end{aligned}
\end{equation}
{where $x_0$, and $x_j^{(0)}$, $j=1,\dots,N$ are arbitrary real numbers.}\footnotemark \\
With these phases, the $N$-soliton solution is then given by the formula 
\be
u(x,t)= 2\pa_x^2 \ln {\bs \tau}(x,t)
\ee
where
\be
\label{tauKdV}
{\bs \tau}(x,t):= {\rm e}^{- \frac{\zeta({\varpi_3})}{{8}{\varpi_3}}x^2} \det \le[\1_N+\mathbb G\ri] \theta_3\le(\frac {x-x_0}{4i\varpi_3} -\mathcal A\ri), 
\ee
$\mathcal A = \frac 1 2 \sum_{j=1}^N (\beta_j - \beta_j^\star)$,
and $\mathbb G$ is  found in Theorem \ref{main}.
\et
\footnotetext{Comparing \eqref{XwithPhases} to \eqref{taugen} we have $\mathbf{X}_0 = \mathbf{X}(0,0).$  }
The formula \eqref{tauKdV} is clearly reminiscent of the Kay-Moses formula for $N$ solitons \cite{KayMoses}.

\br
The arbitrary shifts $x_j^{(0)}$ and $x_0$ in Theorem \ref{solitons!} are obtained by  starting with a vector $\bs u_\varepsilon$ which is fine--tuned so that, for a fixed $\bs \varphi \in\R^N$ we have 
\be
\mathbb B_\varepsilon ({\bs u}_\varepsilon-\bs 1)= i\bs \varphi. 
\ee
Now take the limit $\varepsilon \to 0$;  since the diagonal of $\mathbb B_\varepsilon$ diverges, the vector $\bs u_\varepsilon$ is a suitable perturbation of $\1$. In other words, in the limit the solution ``explores'' a small slice of the Jacobian around the half period. The chosen $\varphi_\ell \in \R$ have the effect of  re-scaling the norming constants by an arbitrary positive constant ${\rm e}^{\varphi_\ell}$. 
\er


\section{Velocity of a single soliton on elliptic background: Bright-and-forward versus dim-and-retrograde}
\label{sect-speed}

Consider here just one hot soliton $b\in (-\infty, e_3)$; as Fig. \ref{bright} shows, we see a hump of taller oscillation on the cnoidal background that moves to the right with velocity $V_j=-E_j/P_j>0$. 
However, for a cool soliton  $c\in (e_2,e_1)$ the disturbance over the cnoidal wave is first of all a {\it subtraction} (i.e. a ``dim soliton''). Moreover, which is more interesting, this disturbance moves to the {\it left} with velocity $V_j = - E_j/P_j<0$. 
The phenomenon was observed in \cite{CERT} using a Whitham approach, and experimentally in \cite{Pelin-F}. Here we observe the same phenomenon by inspecting the explicit solution formula given in Theorem~\ref{thm:1.1}.
\br We can justify the shape of a dim soliton as follows.
With the normalizations used in this paper, the potential $u(x,t)$ can be written as 
\be
\label{trace}
u(x,t)=\frac 1 4\le(2\sum_{j=1}^{N+1}\mu_j -  \sum_{j=1}^{2N+1}E_j\ri)
\ee
where $\mu_j = \mu_j(x,t)$ are the Dirichlet eigenvalues and are one in each of the gaps and $E_j$ are all the branchpoints. Under the degeneration, branchpoints come together in pairs. Consider now the case of a single cool soliton with eigenvalue at $c\in [e_2,e_1]$. Then the trace formula \eqref{trace} implies that (recall that $e_1+e_2+e_3=0$) the value of $u$ is given by  $\frac 1 2(\mu_1 + \mu_2-c)$ where $\mu_1\in [c,e_1]$ and $\mu_2\in [e_2,c]$. This gives the four critical values
\begin{center}
\begin{tabular}{C L L} \toprule 
u(x,t) & \mu_1 & \mu_2 \\  \midrule
\frac {e_1}2 & e_1 & c \\[5pt]
\frac {e_1+e_2-c}2 & e_1 & e_2\\[5pt]
\frac {c}2 & c & c \\[5pt]
\frac {e_2}2 & c & e_2 \\ \bottomrule
\end{tabular}
\end{center}
These four values are the maximum and minimum of the cnoidal wave, and the two critical values of the ``dent'' in the envelope.\hfill $\triangle$
\er
To prove the phenomenon of  on a mathematical level we need to analyze the formula for the soliton;
specializing Theorem \ref{solitons!} to $N=1$ we get
\begin{equation}
\begin{gathered}
\label{singlesoliton}
u(x,t;\beta_1) = 2\pa_x^2\ln {\bs \tau}(x,t;\beta_1) \\
{\bs \tau}(x,t;\beta_1){\rm e}^{\frac{ \zeta({\varpi_3})x^2}{{8}{\varpi_3}}} = \theta_3\le(\frac{x-x_0}{{4}i{\varpi_3}} -\frac {\beta_1-\beta_1^\star}2  \ri) + {\rm e}^{i\le( (x-x_1^{(0)}) P(\beta_1) + t E(\beta_1)\ri)} \theta_3\le(\frac{x-x_0}{{4}i{\varpi_3}} +\frac {\beta_1-\beta_1^\star}2  \ri) 
\end{gathered}
\end{equation}
where $P(\beta_0)=i|P_0|,\  E(\beta_0)$ 
are as in \eqref{Pjs}, \eqref{Ejs} {and we set $V_1 = -E(\beta_1)/P(\beta_1) \in \R$}. 
Rewriting the above in clear form we find 
\be
u(x,t)= - \frac{\zeta({\varpi_3})}{{4}{\varpi_3}} + 2 \frac {\d^2}{\d x^2}\ln\le[
\theta_3 \le(\frac{x-x_0}{{4}i{\varpi_3}} -\frac {\beta_1-\beta_1^\star}2  \ri) + {\rm e}^{-|P_0|(x-x_1^{(0)}- t V_1) }  \theta_3\le(\frac{x-x_0}{{4}i{\varpi_3}} +\frac {\beta_1-\beta_1^\star}2  \ri) \ri].
\ee	
Since both $\theta_3$'s in the logarithm are periodic (and positive) functions of $x$, if 
${x-V_1 t \gg 0}$ 
then the second term is suppressed and the solution looks like the usual cnoidal wave with a shift, while, for 
{$x-V_1t\ll 0$} 
the second term dominates and we can discard the first $\theta_3$ function, so that the solution looks like a cnoidal wave with the opposite shift. 
Thus the only ``disturbance'' occurs when the two addenda are of the same magnitude, namely along the zero curve of the phase ${x-V_1 t}$, 
which is a disturbance with constant negative velocity {for a cool soliton}.

Thus,  we have established that the ``solitons'', whether hot or cool, travel with asymptotic speeds $V(\beta)$ given by 
\be
\label{TrackerVelocity}
V(\beta) = -\frac {E}{P} = \frac 12\frac {\wp'(2{\varpi_3}\beta)}{ \zeta(2{\varpi_3}\beta) - 2 \beta \zeta({\varpi_3}) + \chi \frac {i \pi}{{2\varpi_3}} },\ \ \ 
\beta \in \le(0,\frac 1 2\ri) + \chi\frac \tau 2,
\ee
where $\chi=0$ for hot solitons and $\chi=1$ for cool ones.

Note also that the background cnoidal wave undergoes a shift from the left to the right of the disturbance, clearly visible in the different arguments of the two theta functions.

\begin{figure}
\centerline{\includegraphics[width=0.4\textwidth]{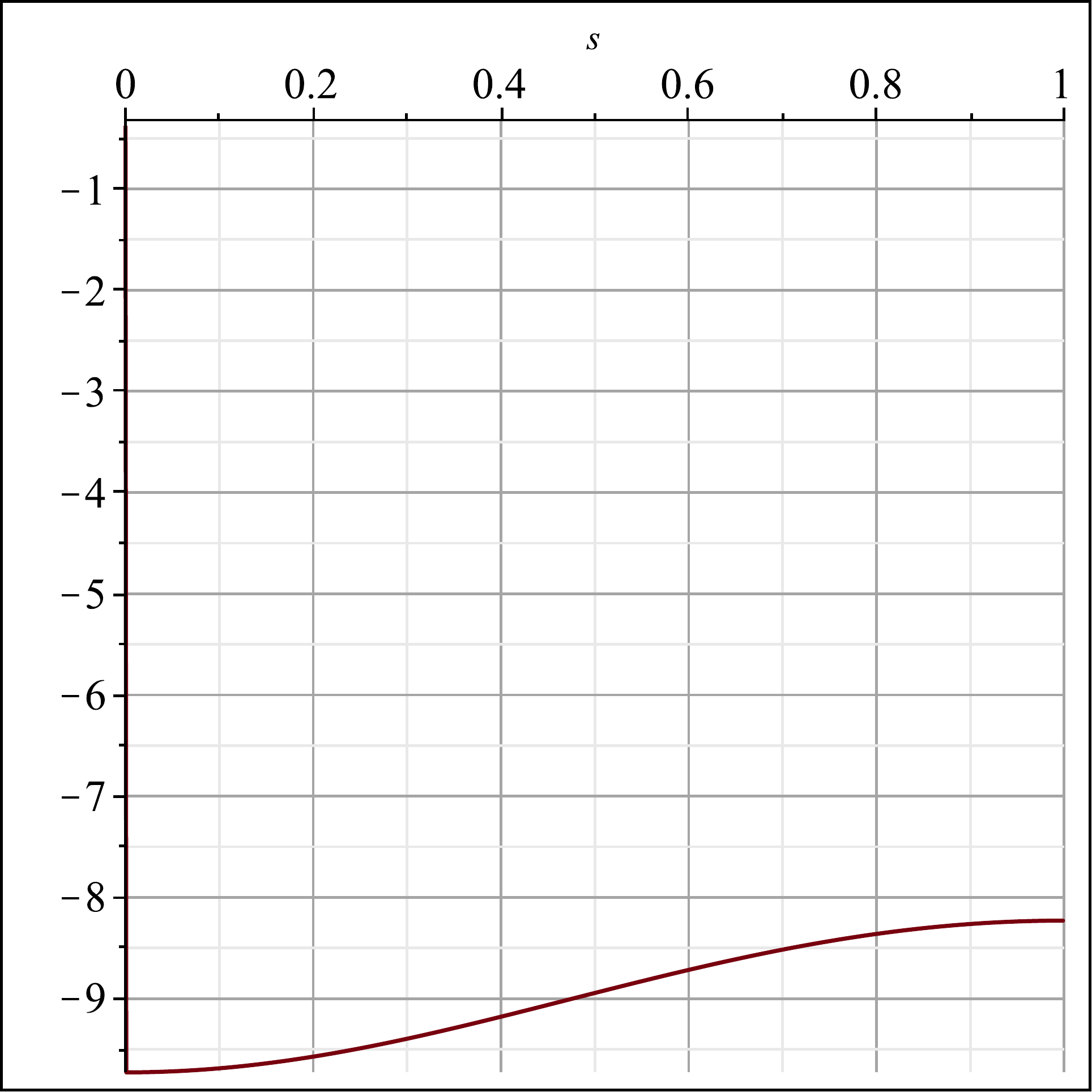}}
\caption{The group velocity $V(\frac \tau 2 +\frac s2)$ \eqref{TrackerVelocity} of a dim soliton as a function of $s$ in $\beta =\frac s 2  + \frac \tau 2 $. Here $\tau \simeq  1.36007\, i$, $e_1 =2$, $e_2 \simeq 1$, $e_3=-3$, 
${\varpi_3} \simeq  -0.742206 i$, $\varpi_2 \simeq 1.009452$. }
\label{groupvelfig}
\end{figure}

The plot of the speed is in Fig. \ref{groupvelfig}, while in Figures \ref{dim}, \ref{bright}, \ref{dimbright} we show, respectively, a single dim soliton (retrograde), a single bright soliton (moving right) and the two-soliton solution obtained from Theorem \ref{solitons!}.

\section{Scattering of soliton pairs over cnoidal background}
\label{scattering}

First, let us verify that very energetic solitons have the expected velocity as an ordinary soliton, which follows from the dispersion relation for the phase
\be
\vartheta_{free} = x\sqrt{|b|} - t |b|^\frac 32
\ee
to be $V_{free}(b)=|b|$, which can also be seen from the single soliton formula \eqref{singlesolitongenus0}.
The limit $b \to -\infty$ corresponds, in the Jacobian, to $\beta \to 0^+$; using the Laurent expansion of $\zeta(s) = \frac 1 s + \mathcal O(s)$ and $\wp'(s) = -\frac 2{s^3} + \mathcal O(s)$ we find
\be
V(\beta) \sim -\frac 1{{4}{\varpi_3}^2\beta^2} \sim -\wp(2{\varpi_3}\beta) =-b = V_{free}(b).
\ee 
Before proceeding we need to define what we mean by the ``position of the soliton" since clearly (see Figures \ref{dim}, \ref{bright}, \ref{dimbright}) the ``solitons'' over the background are not coherent structures  but rather (more or less)  {\it localized} {disturbances} of the cnoidal background. This difficulty is in contrast with the standard case of solitons on a zero background. 

To come up with a working definition we look again at the expression for the single soliton \eqref{singlesoliton}: we define the {\it instantaneous} location as the position, relative to the \textit{ballistic} motion $x =V t =  -E/P t$, where the two addenda in the rhs of \eqref{singlesoliton} are equal. Specifically,  setting $V (\beta)= -\frac {E(\beta)}{P(\beta)} \in \R$, and $x(t) =  x_1^{(0)}+ V(\beta) t + \Phi(t)$ equating the addenda in \eqref{singlesoliton} gives:
\begin{multline}
{\rm e}^{-|P(\beta)| \Phi(t) }C  \theta_3\le(\frac{V(\beta)t + \Phi(t)}{4i{\varpi_3}} +\frac {\beta-\beta^\star}2  \ri)=\theta_3\le(\frac{V(\beta)t + \Phi(t)}{4i{\varpi_3}} -\frac {\beta-\beta^\star}2  \ri) 
 \Rightarrow \ \ \
 \\
 \label{instantphi}
 \Phi(t) = - \frac 1{|P(\beta)|} \ln 
 \frac{\theta_3\le(\frac{V(\beta)t + \Phi(t)}{4i{\varpi_3}} -\frac {\beta-\beta^\star}2  \ri) }{\theta_3\le(\frac{V(\beta)t + \Phi(t)}{4i{\varpi_3}} +\frac {\beta-\beta^\star}2  \ri)}  
 {+ \frac 1{|P(\beta)|} \ln C},
\end{multline}
where we note that $P(\beta) = i|P(\beta)|$.
\bl\label{lem-Phi}
For any $\beta \in(0,\hf)+ \{0,\frac \tau 2\}$ equation \eqref{instantphi} has a unique solution $ \Phi(t)$. Moreover,  $\Phi(t)$ is continuous and has period $\frac{4i\varpi_3}{V(\beta)}$.
\el
\noindent {\bf Proof.}
The equation \eqref{instantphi} defines a unique function $\Phi(t)$ via the implicit function theorem. To see it we observe that the r.h.s. is a periodic, bounded function  $R(\Phi,t)$ of $\Phi$  so that it intersects the line $L(\Phi)=\Phi$. To see uniqueness it is sufficient to show that the function $F(\Phi,t):= L(\Phi)-R(\Phi,t)$ is a monotonic (increasing, as it turns out) function of $\Phi$. 
Now, we have
\be
\frac \pa{\pa \Phi} R(\Phi,t) 
= \frac {-1}{4i\varpi_3|P(\beta)|}\frac{\theta_3'\le(\frac {V(\beta)t + \Phi}{4i\varpi_3} -s \ri)}{\theta_3\le(\frac {V(\beta)t + \Phi}{4i\varpi_3} -s \ri)}\Bigg|_{s=\frac{\beta-\beta^\star}2}^{s=\frac{\beta^\star-\beta}2}
= \frac {-1}{4i\varpi_3 |P(\beta)|}\frac{\theta_1'\le(\frac {V(\beta)t + \Phi}{4i\varpi_3} + \frac{\tau+1}2 -s \ri)}{\theta_1\le(\frac {V(\beta)t + \Phi}{4i\varpi_3} + \frac{\tau+1}2-s \ri)}\Bigg|_{s=\frac{\beta-\beta^\star}2}^{s=\frac{\beta^\star-\beta}2}. 
\ee 
We are going to show that 
\be
\frac \pa{\pa \Phi} R(\Phi,t) \leq 1 
\label{dR<1}
\ee
which is sufficient as long as the equality holds only at isolated points.
Indeed we have (recalling $\varpi_3\in i\R_-$) 
\be
\max_{d\in\R} \frac \pa{\pa \Phi} R(\Phi,t) 
=\frac {-1}{4|\varpi_3| |P(\beta)|}
\min_{d\in[0,1]}
\frac{\theta_1'\le (s+\frac {\tau+1}2+d  \ri)}{\theta_1\le(s+\frac {\tau+1}2+d \ri)}\Bigg|_{s=\frac{\beta-\beta^\star}2}^{s=\frac{\beta^\star-\beta}2}.
\ee
Recall now from \eqref{Pjalt}  that $P(\beta) = \frac 1{4\varpi_3} \frac {\theta_1'(s)}{\theta_1(s)}\bigg|_{s=\beta^\star}^\beta$ so that 
\be
\max_{d\in \R} \frac \pa{\pa \Phi} R(\Phi,t) =
\frac{-1}{
\frac {\theta_1'(s)}{\theta_1(s)}\bigg|_{s=\beta^\star}^\beta
}
\min_{d\in[0,1]}\le(
\frac{\theta_1'\le (s+\frac {\tau+1}2+d  \ri)}{\theta_1\le(s+\frac {\tau+1}2+d \ri)}\Bigg|_{s=\frac{\beta-\beta^\star}2}^{s=\frac{\beta^\star-\beta}2}\ri). \label{iii}
\ee
We now show that the minimum  is achieved at $d=0$; the minimum is $-1$ for cool solitons, and strictly larger for hot solitons,  (thus proving \eqref{dR<1}) and it is an (isolated) critical point. 
Let us denote $\rho:= \frac {\beta^\star - \beta}2\in [0,1/2]$ and note that it is always real.
Consider the function 
\be
F(d):= \frac{\theta_1'\le (s+\frac {\tau+1}2+d  \ri)}{\theta_1\le(s+\frac {\tau+1}2+d \ri)}\Bigg|_{s=-\rho}^{s=\rho}.
\ee
This is an elliptic function of $d$, real valued for $d\in \R \cup \R+\frac \tau 2$.
Its poles are at $d = \pm \rho + \frac {\tau+1}2$ and its zeros (by Abel's theorem) are of the form $\pm \varepsilon + \frac 1 2\in (0,1)$; indeed we find $F(0)<0$ and $F(\frac 1 2)>0$. Moreover the only critical points on the real axis are $d=0,\frac 1 2$. The fact that $d=0$ is a minimum is easily verified by the second derivative test noting that 
\begin{gather}
F'(0) = 4\varpi_3^2\le[\wp\le(2\varpi_2\le(\frac {\tau+1}2 -\rho\ri) \ri) - \wp\le(
2\varpi_3\le(\frac {\tau+1}2 +\rho\ri) \ri)\ri]=0\\
F''(0) = 16\varpi_3^3 \wp'\le(2\varpi_3\le(\frac{\tau+1}2-\rho\ri)\ri)>0
\end{gather}
where the last inequality is due to the fact that the points $\frac {\tau+1}2-\rho$ correspond to the points in the gap $[e_1,e_2]$ on the first sheet, where $\wp'\in i\R_-$, see \eqref{2.5}. Since $F$ is periodic and has only one other critical point at $d=\frac 12$ (which must be a maximum) we have shown that 
\be
\label{F>0}
F(d)\geq F(0), \ \ \forall d\in \R.
\ee 
\\[5pt]
Consider the case of a cool soliton $\beta  = \rho + \frac \tau 2$, $\beta^\star = \frac {\tau}2+1-\rho$. Observing that $\frac{ \tau+1} 2 + \frac {\beta^\star-\beta}2 = \beta^\star$ and 
 $\frac{ \tau+1} 2 - \frac {\beta^\star-\beta}2 = \beta$,
 we see that at the minimal value $d=0$ the right hand side of \eqref{iii} is 1. 
  Then  formula \eqref{iii}  gives exactly \eqref{dR<1}, with the equality being achieved for $d=0$. 
\\[5pt]
Consider now the case of a hot soliton $\beta \in [0,\frac 1 2]$. In view of \eqref{iii}, to prove \eqref{dR<1}
we now have to show that 
\be\label{hot-ineq}
		\left|\pa_\rho \ln{\theta_1\le(\rho-\hf\ri)}\right| \geq 
	\left|\Re\pa_\rho \ln{\theta_1\le(\rho-\frac{\tau+1}2\ri)}\right |, 
	\ee
where $\rho=\frac{\beta^\star-\beta}{2}\in [0,\hf)$.  Using \eqref{2.67} reduces \eqref{hot-ineq} to
\be\label{hot-ineq1}
|\Im\zeta(\varpi_1+\varpi_3-\trho)|\leq |\Im
\zeta(\varpi_3-\trho)|
\ee
where $\trho=2\varpi_3\rho$.
The value  $\trho=0$ turns \eqref{hot-ineq1} into an equation. Thus, 
to prove \eqref{hot-ineq} it is sufficient to show that 
\be\label{hot-ineq2}
|\wp(\varpi_1+\varpi_3-\trho)|\leq 
|\wp(\varpi_3-\trho)|
\ee
for every $\trho \in[0,\varpi_3]$.
That follows from the identities (\cite{Hancock})
\be \label{iden-1}
\wp(\varpi_1+\varpi_3-\trho)=e_2+\frac{(e_2-e_1)(e_2-e_3)}{\wp(\trho)-e_2},\\
\wp(\varpi_3-\trho)=e_3+\frac{(e_3-e_1)(e_3-e_2)}{\wp(\trho)-e_3}.
\label{iden-2}
\ee
Since both terms in \eqref{iden-2} are negative and $|e_3|>|e_2|$, the desired result follows from 
\be
\left|\frac{(e_2-e_1)(e_2-e_3)}{\wp(\trho)-e_2}\right|\leq
\left|\frac{(e_3-e_1)(e_3-e_2)}{\wp(\trho)-e_3}\right|,
\ee
which is clearly true for any $\trho \in[0,\varpi_3]$.

Once existence and uniqueness of solution $\Phi(t)$ of  \eqref{instantphi} is established, the contimuity of  $\Phi(t)$ (in fact, smoothness) follows from the 
Implicit Function Theorem.   Since $\theta$ in \eqref{instantphi} is a period 1 function, it is obvious that if $\Phi(t_0)$ solves \eqref{instantphi}  with $t=t_0$, it also solves \eqref{instantphi}  with $t=t_0+ \frac{4i\varpi_3}{V(\beta)}$.  Now the uniqueness of solution of \eqref{instantphi} implies that $\Phi(t)=\Phi(t+ \frac{4i\varpi_3}{V(\beta)})$.
\QED

	It follows from \eqref{instantphi} and Lemma \ref{lem-Phi} that the average of $\Phi(t)$ over a period is  {$\frac 1{|P(\beta)|}\ln \mathcal{K}$}, because  {($\Phi(t)$ is the difference of two shifted logarithms of $\theta$ with the same period plus a constant)}:
\be
\Phi(t) &= - \frac 1{|P(\beta)|} \ln 
 \frac{\theta\le(\frac{V(\beta)t + \Phi(t)}{4i{\varpi_3}} -\frac {\beta-\beta^\star}2  \ri) }{\theta\le(\frac{V(\beta)t + \Phi(t)}{4i{\varpi_3}} +\frac {\beta-\beta^\star}2  \ri)}
 +\frac{ \ln \mathcal{K}}{|P(\beta)|} \ \Rightarrow \ \ \ \langle \Phi \rangle = \frac 1 T \int_0^T \Phi(t)\d t= \frac{ \ln \mathcal{K}}{|P(\beta)|},
\ee
where $T = \frac {4i\varpi_3}{V(\beta)}$ is the period.
Now, suppose that two solitons  corresponding to the points $\beta_1, \beta_2\in 
\le(0 , \frac 1 2 \ri) + \{0,1\}\frac \tau 2$ are localized at $t\to \pm \infty$ around the positions
\be
x_1(t;\beta_1,\beta_2) &= t V(\beta_1) + \Phi^{(\pm)}_1(t,\beta_1,\beta_2),\ \ \ t \to \pm \infty
\\
x_2(t;\beta_1,\beta_2) &= t V(\beta_2) + \Phi^{(\pm)}_2(t,\beta_1,\beta_2),\ \ \ t \to \pm \infty.
\ee
We are not interested in the value of $\Phi^{(\pm)}_{j}$ directly (since the initial position can be changed arbitrarily by a re-definition of the norming constants) but 
in their average difference:
\be
\Delta_j(\beta_1,\beta_2):= \le\langle \Phi^{(+)}_j(t;\beta_1,\beta_2)\ri \rangle -\le\langle \Phi^{(-)}_j(t;\beta_1,\beta_2)\ri\rangle,  \qquad j=1,2.
\ee
  
\bt
\label{scatteringthm}
{\bf [1]}
The deviation from the ballistic trajectory  for a two--soliton interaction is 
\be\label{Delta-beta}
\Delta_1(\beta_1,\beta_2)
 =
 \frac 2 {|P_1|} \ln\le|\frac { 
 \theta_1(\beta_1-\beta^\star_2)
}{
\theta_1(\beta_1-\beta_2)
}\ri|, \qquad 
\Delta_2(\beta_1,\beta_2)
=-
 \frac 2 {|P_2|} \ln\le|\frac { 
 \theta_1(\beta_1-\beta^\star_2)
}{
\theta_1(\beta_1-\beta_2)
}\ri|,
\ee
where $P_j$, $j=1,2$,   are given in Lemma \ref{lem-PE} and the solitons are ordered so that  $V_2<V_1$ (i.e. the soliton number has the larger velocity {(note that the $V_j$'s may be positive or negative, so that larger velocity does not mean larger speed!)}.\\
{\bf [2]} {(
``Conveyer-belt effect"). The background cnoidal wave undergoes an addition of a  shift by $\beta_j-\beta_j^\star$ in its phase  passing from the right to the left of the $j$-th solitary disturbance:
\be
u(x,t) \simeq  \le\{
\begin{array}{lll}
\ds 2 \pa_x^2\ln \theta_3\le(\frac {x} {4i{\varpi_3}} -\mathcal A    \ri)
 & V_1 t <\!\!< x, & \\
\ds 2 \pa_x^2\ln \theta_3\le(\frac {x} {4i{\varpi_3}} -\mathcal A  + \beta_1-\beta_1^\star   \ri)
 & V_2 t <\!\!< x<\!\!< V_1 t,&  t\to +\infty \\
\ds 2 \pa_x^2\ln \theta_3\le(\frac {x} {4i{\varpi_3}} +\mathcal A    \ri)
 & x<\!\!< V_2 t, &   \\[18pt]
 \hline\\
\ds 2 \pa_x^2\ln \theta_3\le(\frac {x} {4i{\varpi_3}} -\mathcal A    \ri)
 & V_2 t <\!\!< x, &    \\
\ds 2 \pa_x^2\ln \theta_3\le(\frac {x} {4i{\varpi_3}} -\mathcal A  + \beta_1-\beta_1^\star   \ri)
 & V_1 t <\!\!< x<\!\!< V_2 t, &  t\to -\infty \\
\ds 2 \pa_x^2\ln \theta_3\le(\frac {x} {4i{\varpi_3}} +\mathcal A    \ri)
 & x<\!\!< V_1 t, &   \\
\end{array}
\ri.
\\
\mathcal A:= \frac 1 2 \sum_{j=1}^2 (\beta_j-\beta_j^\star).
\ee}
\et
Visualizations of the results of theorem~\ref{scatteringthm} are shown in figures~\ref{2dimshift}-\ref{2dimshiftreference}. We call the second effect a ``conveyer-belt'' effect because the solitary disturbance is sort of ``kicking'' the background like a runner on a conveyer belt. A similar shifting effect on the background cnodial wave was previously observed (\cite[Prop. 6.1]{GGJMM22}.) in the context of the modified KdV equation for a single ``trial" soliton of higher velocity passing through an elliptic (genus 1) soliton condensate (also known as condensate limit of a soliton gas, see \cite{ElTo2020}, \cite{El-Rev}).  
\\[3pt]
\noindent {\bf Proof.}
{\bf [1]}
Consider the two solitons, each of which could be hot or cool. 
We denote for brevity $P_j = P(\beta_j), E_j= E(\beta_j), \ V_j = V(\beta_j)$, 
with the velocity $V$ given in \eqref{TrackerVelocity}.

The core of the computation is to analyze the dominant term in the tau-function 
\eqref{tauKdV}, recalling that we need to take the second logarithmic derivative in $x$. 

For this reason it should be evident that the position of the disturbances (solitons) is determined by the asymptotic behaviour of the $2\times 2$ determinant $\det [\1_2 + \mathbb G]$. 
Let us denote, for brevity 
\be
\lambda_j :={\rm e}^{2i\pi \psi_j}=  {\rm e}^{ i (x-x_j^{(0)})  P_j + i t E_j} = {\rm e}^{-
{| P_j|} \le(x-x_j^{(0)} - V_j t\ri)} \in \R,
\ee
where we  have used  that $P_j$ is always in $i\R_+$ (see \eqref{Pjs}), while $E_j\in i\R_-$ for hot solitons and $E_j\in i\R_+$ for cool ones \eqref{Ejs}.

Let us consider for definiteness the hot-on-hot interaction and order the solitons so that $V_1>V_2$. 
As $t\to -\infty$ we consider the location around $x= x_{j}(t) := x_j^{(0)} -V_j|t|$ 
\be
\l_1\simeq 1,\ \ \ \l_2 = {\rm e}^{|P_2| (V_1-V_2)|t| +|P_2|(x_2^{(0)}-x_1^{(0)})  } \gg 1.
\ee
Viceversa, around $x=x_2(t)$ we have that $\l_1$ is exponentially small
\be
\l_1\simeq {\rm e}^{-|P_1|(V_1-V_2)|t| + |P_1|( x_1^{(0)}-x_2^{(0)}) }
\ll 1,\ \ \ \l_2 \simeq 1.
\ee
The general structure of the determinant in \eqref{Noice} when considered in our situation, is 
\be
\det\le[\1_2 + \mathbb G\ri] =\det \le[
\begin{array}{cc}
1 + A\l_1  &  B \sqrt{\l_1\l_2}\\
B \sqrt{\l_1\l_2} & 1+C \l_2
\end{array}
\ri]
\ee
where $A, B, C$ are some periodic, bounded functions of $x$ {\it only}. 

Let us follow the slower of the two solitons, $V_2<V_1$; around $x_2(t)$ (as $t\to -\infty$) $\l_1$ is exponentially small and thus clearly
\be
\det\le[\1_2 + \mathbb G\ri] \simeq 1 + C\l_2,
\ee
and then we have, in this regime (with $\beta = \frac {x}{4i\varpi_3}$)
\be
\theta_3\le(\beta - \mathcal A\ri) \det[\1_2 + \mathbb G] \simeq
\theta_3\le(\beta - \mathcal A\ri) \le(1 + \mathbb G_{22}\ri)
=
\\=
\le (\theta_3\le(\beta- \mathcal A\ri) + {\rm e}^{-|P_2|(x-x_2^{(0)}-V_2 t) } 
 \theta_3\le(\beta - \mathcal A + \beta_2 - \beta_2^\star \ri)
 \le| 
\frac {\theta_1(\beta_1 - \beta_2^\star)}{\theta_1(\beta_1-\beta_2)
}\ri|
\ri)
\label{418}
\ee
and the soliton is located where the two addenda are approximately equal. Letting $x = x_2(t) + \Phi_2^{(-)}$ we have that the effect is maximal at 
\be
\label{-inf}
-|P_2| \Phi_2^{(-)} = \ln\frac {\theta_3\le(\frac {x_2(t) + \Phi_2^{(-)} }{2i{\varpi_3}}  - \mathcal A\ri)}{ \theta_3\le(\frac {x_2(t) + \Phi_2^{(-)} }{2i{\varpi_3}}+ \frac {\beta_2 -\beta_1+ \beta_1^\star- \beta_2^\star}2  \ri)
 \le| 
\frac {\theta_1(\beta_1 - \beta_2^\star)}{\theta_1(\beta_1-\beta_2)
}\ri|}.
\ee
Note that here there is no such thing as an ``exact position'' (even asymptotically) of the soliton disturbance on the background. However the above equation \eqref{-inf} for $\Phi_2^{(-)}$ defines clearly a periodic function of $t$ and the average on a period yields the equation
\be
 \le\langle  \Phi_2^{(-)} \ri\rangle=   \frac 1{|P_2|} \ln   \le| 
\frac {\theta_1(\beta_1 - \beta_2^\star)}{\theta_1(\beta_1-\beta_2)
}\ri|.
\ee
Now consider $t\to +\infty$; in this case, following $x_2(t)$ again we have 
\be
\l_1 \simeq {\rm e}^{|P_1|(V_1-V_2)t + |P_1|(x_1^{(0)}-x_2^{(0)})} \gg 1, \qquad \l_2 \simeq 1.
\ee
Thus the computation of the determinant needs to be done differently; using row operations we obtain 
\be
\det\le[\1_2 + \mathbb G\ri]  
=&(1+ A\l_1) \le( 1+C \l_2-\frac{ B^2 \l_1\l_2}{1+A\l_1}\ri),
\ee
which is nothing but Schur complement formula. Recalling now that $\l_1$ is exponentially large, we have 
\be
\ln \det[\1_2+ \mathbb G] \sim
 \ln \l_1 + \ln \le( 1+C \l_2-\frac{ B^2 \l_2}{A}\ri) + \mathcal O(\l_1^{-1})
=
 \ln \l_1 + \ln \le( 1+\frac{AC - B^2 }{A}\l_2\ri) + \mathcal O(\l_1^{-1})
\ee
Since $\ln \l_1$ is linear in $x$, the solution (which requires two derivatives in $x$) is not affected and we need only consider the middle logarithm.
So effectively we need to analyze 
\be
\theta_3\le( \beta - \mathcal A\ri) \frac{\det[\1_2 + \mathbb G] }{1 + \mathbb G_{11}}
\simeq
\theta_3\le( \beta - \mathcal A\ri) \le(1 + \frac {\det \mathbb G}{\mathbb G_{11}}\ri).
\ee
Using Fay's identity \eqref{Fay} we get 
\be
\det \mathbb G = C_1 C_2 \l_1\l_2
\frac {\theta_3 (\beta+\mathcal A)
\theta_1(\beta_1-\beta_2)\theta_1(\beta_2^\star - \beta_1^\star)}{
\theta_{3}(\beta-\mathcal A)\theta_1(\beta_1-\beta_1^\star)\theta_1(\beta_2-\beta_2^\star)\theta_1(\beta_1-\beta_2^\star)\theta_1(\beta_2-\beta_1^\star)}
\ee
so that \be
\theta_3\le( \beta - \mathcal A\ri) \frac{\det[\1_2 + \mathbb G] }{1 + \mathbb G_{11}}
\simeq
 \theta_3\le( \beta - \mathcal A\ri) + {\rm e}^{-|P_2|(x -x_2^{(0)} - V_2 t) } \frac 
{\theta_3\le( \beta - \mathcal A\ri)\theta_3 ( \beta +\mathcal A)
}
{\theta_3(\beta_1 - \beta_1^\star + \beta - \mathcal A)}
\frac{|\theta_1(\beta_1-\beta_2)|}{|\theta_1(\beta_2-\beta_1^\star)|}.
\ee
Now we set $x = x_2(t) + \Phi_2^{(+)}$ and obtain 
\be
\label{+inf}
-|P_2|\Phi_2^{(+)} =  \ln\frac {{\theta_3\le(\frac {V_2t + \Phi_2^{(-)} }{2i{\varpi_3}}+
\frac{\beta_1 -\beta_2- \beta_1^\star +\beta_2^\star}2 \ri)}
}{
{\theta_3 \le(\frac {V_2t + \Phi_2^{(-)} }{2i{\varpi_3}}+\mathcal A\ri)
}
\frac{|\theta_1(\beta_1-\beta_2)|}{|\theta_1(\beta_2-\beta_1^\star)|}
}.
\ee
Once again, the period average of the dislocation is 
\be
\le\langle\Phi_2^{(+)} \ri\rangle = \frac 1{|P_2|} \ln\frac{|\theta_1(\beta_1-\beta_2)|}{|\theta_1(\beta_2-\beta_1^\star)|}.
\ee 
Now the shift in $x$ location follows from the ratio of \eqref{+inf} and \eqref{-inf}
\be
\frac {\l_2^{(+)}}{\l_2^{(-)}} = {\frac {AC}{AC-B^2}}
\ \ \ \ \Rightarrow \ \ \ \ 
\Delta_{21} = -\frac 1{ |P_2|} \ln\frac {AC}{AC-B^2}.
\ee
Thus the total shift is 
\be
\Delta_2(\beta_1,\beta_2)  = \le\langle \Phi_2^{(+)}\ri\rangle - \le\langle\Phi_2^{(-)}\ri\rangle = 
\frac 2{|P_2|} \ln\frac{|\theta_1(\beta_1-\beta_2)|}{|\theta_1(\beta_2-\beta_1^\star)|}.
\ee
{\bf [2]}
{
Observe the argument of two functions $\theta_3$ in \eqref{418}: on the right ($ \beta = \frac{x}{4i\varpi_3}, \ x=-(V_2-\varepsilon) |t|$) the exponential is exponentially small and the {\it first} $\theta_3$ dominates, with the phase $\frac{x}{4i\varpi_3} -\mathcal A$. On the left ($x = -(V_2+\varepsilon)|t|$) the exponential  is large and the {\it second} $\theta_3$ dominates with the phase $ \frac{x}{4i\varpi_3}  - \mathcal A + \beta_2-\beta_2^\star$.  Similarly for the other transition.  
}
\QED

\section{Elliptic gas of solitons for KdV}
\label{sect-gas}
We now consider the setup of a growing number of degenerating bands  on the elliptic background. Unlike the rest of the paper,  the calculations in this section are mostly formal.
The total number  $N$ of collapsing bands (solitons) of our Riemann surface  $\Rscr_{N+1}(\eps)$  is now linked with the size $2\delta_j>0$ of the $j^\mathrm{th}$ collapsing band  by $2\delta_j=e^{-N\nu_j}$, where  all $\nu_j\geq 0$. 
In this section, it will be more convenient for us 
(following  \cite{FFM}, \cite{El2003})
to  normalize  the differentials  $\d P$, $\d E$ by $\scr B$-periods, i.e., the $\scr B$-periods of $\d P$, $\d E$ corresponding to  all the shrinking bands are zero. 
 The normalization on the background (resolved) part 
 of $\Rscr_{N+1}(\eps)$   is unchanged.
The quasi-momentum and quasi-energy differentials with the new normalization  will be denoted as $\d\hat P$, $\d\hat E$ respectively.
  In view of Figure \ref{cycles}, one can observe that   $\d \hat P$, $\d \hat E$ are gap-normalized differentials, meaning that their periods over each gap are zero.
The relation between the quasi-energy and quasi-momentum used in the previous part of the paper and the gap-normalized versions used here are
\begin{equation}
	\begin{aligned}
	& \d \hat P = \d P - \sum_{k=1}^N c_k \omega_k,    
	&& \bs{c} = \left[ \begin{array}{@{}c|c@{}}  \mathbb{B}^{-1}\!\!\! & \bs{0}_{N\times1} \end{array} \right] \bs{P},
	\\
	&\d \hat E = \d E -  \sum_{k=1}^N  \gamma_k \omega_k,
	&& \bs{\gamma} = \left[ \begin{array}{@{}c|c@{}}   \mathbb{B}^{-1}\!\!\! & \bs{0}_{N\times1} \end{array} \right] \bs{E}.
	\end{aligned}
\end{equation}
Here $\omega_j,\ j=1,\dots, N+1,$ are the normalized holomorphic differentials on $\Rscr_{N+1}(\eps)$, $\mathbb{B}$ is the $N\times N$ block matrix of their $\scr B$-periods from \eqref{Omega.blocks}, and $\bs{P}$ and $\bs{E}$ are the $N+1$ dimensional vectors of $\scr B$-periods of $\d P$ and $\d E$ respectively, \eqref{P/E vectors}.
 According to  \cite{FFM}, \cite{El2003} the 
  $\scr A$-periods of $\d \hat P$, $\d \hat E$, related to the  shrinking bands, define the {\it wavenumbers $k_j$ } and {\it frequences $w_j$}, $j=1,\dots,N$, of the nonlinear wave (finite gap) solutions to the KdV equation defined by $\Rscr_{N+1}(\eps)$.
   Then the Riemann bilinear identities on $\Rscr_{N+1}(\eps)$ yield 
\begin{equation}\label{RBL-bulk}
\begin{gathered}
\lim_{\eps\to 0}\sum_{l=1}^{N}\oint_{\scr A_l}\d \hat P\oint_{\scr B_l}\o_j=\frac{2\pi i}{4{\varpi_3}}\frac{ \o_j}{\d \beta} \Big\vert_{\beta=0}, \\
\lim_{\eps\to 0}\sum_{l=1}^{N}\oint_{\scr A_l}\d \hat 
E \oint_{\scr B_l}\o_j=\frac{2\pi i}{ {32} {\varpi_3}^3}\partial^2_{\beta}\frac{ \o_j}{\d \beta} \Big\vert_{\beta=0}, 
\end{gathered}
\end{equation}
where $j=1,\dots, N$ and 
\begin{equation}\label{RBL-last}
\begin{gathered}
\lim_{\eps\to 0}\sum_{l=1}^{N}\oint_{\scr A_l}\d \hat P\oint_{\scr B_l}\o_{N+1}-\oint_{\scr B_{N+1}}\d \hat P=\ \frac{2\pi i}{4\varpi_3} 
, \\
\lim_{\eps\to 0}\sum_{l=1}^{N}\oint_{\scr A_l}\d \hat E  \oint_{\scr B_l}\o_{N+1}-\oint_{\scr B_{N+1}}\d \hat E =0.
\end{gathered}
\end{equation}
Equations  \eqref{Blm} 
imply that 
\be\label{RBL-back}
k_{N+1}:=\oint_{\scr B_{N+1}}\d \hat P=\sum_{j=1}^N k_j (\beta_j-\beta_j^\star)-{\frac{\pi i}{2{\varpi_3}}} ,    \qquad   w_{N+1}:=\oint_{\scr B_{N+1}}\d \hat E  = \sum_{j=1}^N w_j (\beta_j-\beta_j^\star),
\ee
where $k_j,w_j$ denote the $\mathscr{A}_j$ periods of $\d \hat P,\d\hat E $ respectively, $j=1,\dots,N$. 
Substituting \eqref{Blm} into \eqref{RBL-bulk}, we obtain
\begin{equation}\label{RBL-bulk-1}
\begin{gathered}
O(-\ln \delta_j)k_j+
\sideset{}{'}\sum_{\ell=1}^{N} k_\ell\ln \le|
\frac{\theta_1 (\beta_j - \beta_\ell)}{\theta_1(\beta_j - \beta_\ell^\star)}
\ri |
=
-\frac{i\pi}{2{\varpi_3}}\le[ \le.\partial_\beta\ln \theta_1 (\beta)\ri|_{\beta=\beta_j}+{\pi}i\chi(\beta_j)\ri], \cr
O(-\ln \delta_j)
w_j+\le.
\sideset{}{'}\sum_{\ell=1}^{N} w_\ell\ln \le|
\frac{\theta_1 (\beta_j - \beta_\ell)}{\theta_1(\beta_j - \beta_\ell^\star)}
\ri |
=
- \frac{i\pi}{{16}{\varpi_3}^3}\partial^3_{\beta}\ln \theta_1 (\beta)\ri|_{\beta=\beta_j}, 
\end{gathered}
\end{equation}
where $\chi(\beta)$ is the characteristic function of $(e_2,e_1)$ (the cool solitons),  $ \sum'$ denotes the summation with $\ell\neq j$  and
$j=1,\dots, N$. 
Equations \eqref{RBL-bulk-1} imply that for a fixed $N\in\N$ and  some  $\delta_j\to 0$ we will necessarily have $k_j,w_j\to 0$. Therefore, following \cite{ElTo2020}, we will refer to  the wavenumbers $k_j$ and the frequences $w_j$, $j=1,\dots,N$,  as {\it solitonic}, whereas the
remaining  wavenumber $k_{N+1}$ and the frequency $w_{N+1}$ will be refered to as  {\it carrier}.

The last part of \eqref{2.67} can be conveniently rewritten as 
\be\label{iden-ell}
\pa_\beta \ln\theta_1(\beta)=-4\varpi_3\zeta(\varpi_3)\beta+ 2 \varpi_3 \zeta(2\varpi_3\beta),
\ee
which yields
\be\label{iden-ell'}
\partial_\beta^2\ln \theta_1 (\beta)= -4\varpi_3\zeta(\varpi_3)-  4{\varpi_3}^2\wp (2\varpi_3\beta), \qquad 
\partial_\beta^3\ln \theta_1 (\beta)=
-  8{\varpi_3}^3\wp' (2\varpi_3\beta), 
\ee
where the former expression  essentually is given in  
\cite[Eq. 1035.01]{Byrd}.
Using now  the asymptotics of $\oint_{B_j}\o_j$, $j=1,\dots,N$ from \cite{Ven-89}, \cite{El2003}, \cite{El-Rev} and \eqref{iden-ell}, \eqref{iden-ell'}, we can rewrite \eqref{RBL-bulk-1} as
\begin{equation}\label{RBL-bulk-2}
\begin{gathered}
 N\nu_j
k_j+
\sideset{}{'}\sum_{\ell=1}^{N} k_\ell \ln \le|
\frac{\theta_1 (\beta_j - \beta_\ell)}{\theta_1(\beta_j - \beta_\ell^\star)}
\ri |
=  -{i\pi}\left[\zeta( 2\varpi_3\beta_j)-2\zeta(\varpi_3)\beta_j+\frac{i\pi} {2\varpi_3}\chi(\beta_j)\right], \cr
 N\nu_j
w_j+
\sideset{}{'}\sum_{\ell=1}^{N} w_\ell \ln \le|
\frac{\theta_1 (\beta_j - \beta_\ell)}{\theta_1(\beta_j - \beta_\ell^\star)}
\ri |
=\frac{ i\pi}{2}\wp'( {2{\varpi_3}}\beta_j).
\end{gathered}
\end{equation}

Let us now assume that, as $N\to\infty$, the degenerating bands are accumulating on some interval  (or collection of intervals) $\G\subset \R$ that is separated from the stationary bands of $\Rscr_N$,
i.e., $\G\subset(-\infty,e_3)\cup(e_2,e_1)$.
 Let $\varphi(z)$ be the (limiting)  probability density of the centers $b_j$ of the $j=1,\dots,N$ shrinking bands on $\G$, ${\rm supp} ~\varphi =\G$, and $\nu(z)\geq 0$ be a smooth function on $\G$ interpolating $\nu_j=
 	\nu(b_j)$ on $\G$. Such an $N\to\infty$ limit, subject to some additional restriction, see \cite{El2003}, \cite{ElTo2020},  is called the thermodynamic limit of $\Rscr_N$.  In the thermodynamic limit,  equations  \eqref{RBL-bulk-2} for the solitonic wavenumbers and frequencies become
\begin{equation}\label{NDR-bulk-1} 
\begin{gathered}
\int_{\tilde\G} \ln \le|
\frac{\theta_1 (\eta - \beta)}{\theta_1(\eta+ \beta-1-\chi(\beta))}
\ri |
u(\beta) d\beta + \s(\eta)u(\eta)=  -\frac{i}{  2}\le[
\zeta({2{\varpi_3}}\eta)-2 \zeta(\varpi_3)\eta+\frac{\pi i}{2\varpi_3}\chi(\eta)\ri], \cr
\int_{\tilde\G}  \ln \le|
\frac{\theta_1 (\eta - \beta)}{\theta_1(\eta+ \beta-1-\chi(\beta))}
\ri |
v(\beta) d\beta + \s(\eta)v(\eta)=  \frac{i}{ 4}\wp'({2{\varpi_3}}\eta),
\end{gathered}
\end{equation}
where:  $\tilde\G=\wp(2{\varpi_3}\G)$ is the image of $\G$ in the Jacobian;
\be\label{u,v}
u(\eta)=\frac{\hat u(\eta)\hat \varphi(\eta)}{2\pi}, \qquad v(\eta)=\frac{\hat v(\eta)\hat \varphi(\eta)}{2\pi} \qquad \s(\eta)=\frac{\nu(\wp(2{\varpi_3}\eta))}{\hat \varphi(\eta)}
\ee
with $\hat u(\eta), \hat v(\eta)$ interpolating $Nk_j, Nw_j$ at $\beta_j$, $j=1,\dots, N$ respectively, (compare with \cite{El-Rev}, Section 3.2) and; 
\be
\hat\varphi(\beta)=2{\varpi_3}\varphi(\wp(2{\varpi_3}\beta))\wp'(2{\varpi_3}\beta).
\ee
The expressions for the thermodynamic limit of the carrier  (background) wave-number $\tilde k=\lim_{N\to\infty} k_{N+1}$ and frequency $\tilde w=\lim_{N\to\infty} w_{N+1}$
obtained from \eqref{RBL-back}, are  
\be\label{tilde-u,v}
 \tilde k=2\pi\le[\int_{\tilde\G} (2\beta-1-\tau\chi(\beta))u(\beta)d\beta-\frac{ i}{2\varpi_3}\ri], ~~~
 \tilde w=2\pi \int_{\tilde\G} (2\beta-1-\tau\chi(\beta))v(\beta)d\beta.
\ee

Similarly to the breather gas in focusing NLS setting, see \cite{ElTo2020}, equations \eqref{NDR-bulk-1} form the so-called solitonic  nonlinear dispersion relation (NDR) for the KdV soliton gas on the elliptic background, which we can losely speaking call the elliptic KdV gas.  The carrier NDR is given by \eqref{tilde-u,v}. It is worth mentioning that, like in the case of the fNLS soliton gas (\cite{ElTo2020}), the imaginary part of the general NDR  \eqref{RBL-bulk}-\eqref{RBL-last} form the solitonic NDR, whereas the real part of \eqref{RBL-bulk}-\eqref{RBL-last} form the carrier  NDR.

Following the approach of  \cite{ElTo2020}, Section 5, we derive the equation of states for the speed $s(\eta)= -\frac{v(\eta)}{u(\eta)}$ of  element of the gas (tracer soliton)
\be\label{eq-st}
s(\eta)=s_0(\eta)+ \int_{\tilde\G}\Delta(\eta,\beta)[s(\eta)-s(\beta)]u(\beta)d\beta,
\ee
where 
\be\label{s,Delta}
s_0(\eta)=\frac{\wp'(2{\varpi_3}\eta)}{{ 2} 
	[\zeta(\eta)-2 \zeta(\varpi_3)\eta+\frac{\pi i}{{2{\varpi_3}}}\chi(\eta)]},
\ee
has the meaning of the speed of a single solition on the elliptic background (free speed) and 
\be\label{Delta}
\Delta(\eta,\beta)= \frac {i \ln \le|
	\frac{\theta_1 (\eta - \beta)}{\theta_1(\eta+ \beta-1-\chi(\beta))}
	\ri |^{2}}
{\zeta( {2{\varpi_3}}\eta)-2 \zeta(\varpi_3)\eta+\frac{i\pi}{2\varpi_3}\chi(\eta)},
\ee
has the meaning of the (total) phase shift of  the $\eta$-soliton (with the spectral parameter $\eta$) when it interacts with the $\beta$-soliton. All the solitons are considered on the elliptic background. 
We want to point out that the free speed and the phase shift expressions derived in this section for the  elliptic KdV gas, see \eqref{s,Delta}, \eqref{Delta} respectively, coincide with the corresponding expressions \eqref{TrackerVelocity} and \eqref{Delta-beta}-\eqref{Pjs} respectively, established  in the previous sections of this paper for solitons on the elliptic background. 

The equation \eqref{eq-st} is an integral equation to find $s(\eta)$ considering $u(\eta)$ given.  It was first suggested by V. Zakharov  in \cite{Zakh-71} for diluted soliton gas (with no background) and was later extended by G. El \cite{El2003} to the case of dense gases. 
Equation \eqref{eq-st} naturally extends results of \cite{Zakh-71} , \cite{El2003} to the soliton gas on an elliptic background. 

Finally, let us consider the large $N$ limit of the  background shift $\mathcal A$, see Theorem \ref{thm:1.1}, and the averaged total shift  $S(b)$ of the soliton, parametrized by $b$, given by \eqref{Total-shift}. Here we assume that the 
soliton eigenvalues $b_j$ accumulate  on $\G$ with the limiting probability density $\varphi(z)$. Then in the large $N$ limit we have
\begin{equation}	
\begin{gathered}
		\frac{\mathcal A}N\sim
		\int_{\tilde\G} (\beta-\hf-\tau\chi(\beta))\hat\varphi(\beta)d\beta, 
		 \cr
	S(b)=	 4\Bigg|\varpi_3\frac{\theta_1\le( \beta ;\tau\ri)}{\theta_1'\le( \beta ;\tau\ri)} \Bigg|_{\beta=\eta^\star}^{\beta=\eta}\Bigg|
		\int_{\tilde\G}    \ln \le|
	\frac{\theta_1 (\eta - \beta)}{\theta_1(\eta+ \beta-1-\chi(\beta))}
	\ri | \text{sign}[\wp(2\varpi_3\beta)-\wp(2\varpi_3\eta)]
	\tilde\varphi(\beta) d\beta,
			\end{gathered}
\end{equation}
where $\eta^\star=1+\tau\chi(\eta)-\eta$.
The latter formula can be interpreted as the average total shift of a {\it tracer} soliton of the KdV soliton gas, parametrized by $b$, on the elliptic background.

\appendix
\renewcommand{\theequation}{\Alph{section}.\arabic{equation}}
\section{Arbitrary genus}
\label{genusg}
Suppose $\mathcal C_\varepsilon$ is a family of degenerating curves which, as $\varepsilon\to 0$ becomes an {\it irreducible} nodal curve $\mathcal C_0$ with a resolution $\mathcal C$ of genus $g$.  On the resolution $\mathcal C$ we fix a basis of $\scr A, \scr B$ cycles (in homotopy first, and then as their image in homology) and dissect the surface $\mathcal C$ so as to obtain a simply connected domain $\mathcal L$. We choose the cycles that avoid all the nodes. 

On the resolution there are $N$ pairs of points $p_j, p_j^\star$, $j=1,\dots, N$  which correspond to the two point of the resolution of the nodes. 
The basis of cycles of the family is chosen so that 
\begin{itemize}
\item The {\it vanishing cycles} $\scr A_{g+j}$ reduce in the limit to small counterclockwise circles around $p_j \in \mathcal C_0$;
\item the {\it logarithmic cycles} $\scr B_{g+j}$ reduce to mutually disjoint paths from $p_j$ to $p_j^\star$ within the fundamental dissection $\mathcal L$. 
\end{itemize}

\subsection{Notations and conventions.}
We denote by $\omega_j$ the normalized holomorphic differentials on the resolved nodal curve $\mathcal C$ and by ${\bs \tau}$ the $g\times g$ matrix of normalized $\scr B$--periods:
\be
\oint_{\scr A_j} \omega_k =\delta_{jk}, \qquad {\bs \tau}_{jk} = \oint_{\scr B_j} \omega_k = {\bf \tau}_{kj}.
\ee
We denote by $\A$ the Abel map $\A:\mathcal C \to \mathbb J(\mathcal C)$ in the Jacobian of $\mathcal C$, with basepoint $p_0$:
\be
\A (p)= \int_{p_0}^{p} \le[
\begin{array}{c}
\omega_1\\
\vdots\\
\omega_g
\end{array}\ri]
\ee
where the path is the unique path within $\mathcal L$. 
 We denote by $\mathcal K = \mathcal K_{p_0}$ the vector of Riemann constants \cite{Kra}.
 Finally we have the Riemann theta function 
\be
\Theta({\bs X}; {\bs \tau}) =\sum_{{\bf n} \in \Z^g} \exp i\pi\le(
{\bf n}^\transpose \bs \tau \bs n +2 \bs n^\transpose \bs X
\ri)
\ee
If $\Delta = [\vec \varepsilon, \vec \delta]$ denotes a half--period $\Delta  = \frac 1 2\vec \varepsilon  + \frac 1 2\bs \tau \vec \delta \in \frac 1 2 \Z^g + \frac 1 2 \bs \tau\Z^g$, the {\it Theta function with characteristic $\Delta$} is 
\be
\Theta_{\Delta}({\bs X}; {\bs \tau}) =\sum_{  {\bf n}  \in \Z^g +\frac 1 2  \vec \delta } \exp i\pi\le(
{\bs n}^\transpose \bs \tau \bs n +2 {\bs n}^\transpose \le( \bs X- \frac 1 2\vec \varepsilon\ri) \ri).
\ee
A simple verification shows that (we omit $\bs \tau$ from the notation for brevity) 
\be
\Theta_{\Delta}(-{\bs X}) = (-1)^{\vec \delta \cdot \vec \varepsilon}\Theta_{\Delta}({\bs X}) 
\ee
and thus such half--periods are called {\it even} or {\it odd} according to the parity of $\vec \delta \cdot \vec \varepsilon \in \Z$.
Those odd periods for which the gradient of $\Theta_{\Delta}$ at the origin is not zero are called {\it non-singular}. 
A theorem \cite{Mumford} guarantees that there is at least one odd, non-singular such odd period. From now on we denote by $\Delta$ one such choice.\\[3pt]
The main use of $\Theta_\Delta$ is explained below.
If $p, p^\star$ are two arbitrary points on $\mathcal C$ and we want the expression of the unique third--kind differential with poles at $p$ (with residue $+1$) and at $p^\star$ with residue $-1$  and vanishing $\scr A$--periods is then 
\be
\rho_{p,p^\star}(q)= \d_q \ln \frac{\Theta_\Delta\le(
\A(q)-\A(p)\ri)}{\Theta_\Delta(\A(q)-\A(p^\star))}.
\ee
It is of paramount importance that the contour of integration of the Abel map be taken within the same fundamental domain $\mathcal L$, for otherwise the $\scr A$--periods are not zero. 

Now, following the same logic as explained in the text, we have that the degeneration limit of the $\scr B$ matrix of periods is 
\be
\bs \Omega_\varepsilon  \simeq  \le[
\begin{array}{c|c}
\bs \tau& \bs \mu\\
\hline
\bs\mu^\transpose & \mathbb B
\end{array}
\ri],
\ee
where $\mathbb B\in {\rm Mat}_{N\times N}$, $\bs\tau\in {\rm Mat}_{g\times g}$ and $\bs\mu\in {\rm Mat}_{g\times N}$. The diagonal of $\mathbb B$ diverges to $+i\infty$ while 
\be
\label{higherB}
\mathbb B_{\ell m}\to \frac 1{2i\pi} \ln \le(
\frac {\Theta_\Delta(\A(p_\ell ) - \A(p_m))\Theta_\Delta(\A(p_\ell^\star) - \A(p_m^\star))}{\Theta_\Delta(\A(p_\ell)- \A(p_m^\star))\Theta_\Delta(\A(p_\ell^\star)- \A(p_m))}
\ri)
\\
\mu_{\ell, a}\to \int_{p_\ell^\star}^{p_\ell} \omega_a = \A_a(p_\ell)- \A_a(p^\star_\ell).
\ee
In other words, the $\ell^{\mathrm{th}}$ column of $\bs \mu$ tends to the Abel map on 
the resolved limiting curve of the difference of the points of the resolution of the  $\ell^{\mathrm{th}}$ node. 
The reader may object about the ambiguity on the determination in the formula \eqref{higherB}, to which we refer to the discussion in the main text; we assume, however, that the real part in \eqref{higherB} has been completely determined (by the choice of cycles).
It is now really an exercise to track the same proof of Theorem \ref{main} and obtain 
\bt
Let ${\bf X}\in \C^{g+N}$ be the vector 
\be
{\bf X} = \le[
\begin{array}{c}
\bf w
\\
\bf z
\end{array}
\ri], \ \ {\bf w} \in \C^g,\ \ \ { \bf z}\in \C^N.
\ee
The Riemann--Theta function $\wt \Theta$ of the degenerating family of curves $\mathcal C_\varepsilon$ satisfies
\be
\lim_{\varepsilon \to 0}
{\widetilde{\Theta}}\le( {\bf X} - \frac 1 2 \bs \Omega_\varepsilon \le[
\begin{array}{c}
0\\
\vdots \\
0\\
1\\
\vdots 
\\1
\end{array}\ri];\bs \Omega_\varepsilon \ri)
=
\det\le[\1_N+\mathbb G\ri] \Theta\le(
{\bf z} - \mathcal A; \bs \tau
\ri)
\ee
where 
\be
\mathcal A&= \frac 12 \sum_{j=1}^N \le(\A(p_j)-\A(p_j^\star)\ri)
\\
\mathbb G_{\ell,m} &= 
\frac {\Theta(\A(p_\ell) - \A(p^\star_m) + {\bf w}-\mathcal A)}{
\Theta_\Delta\le(\A(p_\ell)- \A(p^\star_m)\ri)
\Theta\le({\bf w}-\mathcal A\ri)
} C_\ell {\rm e}^{i\pi (\psi_\ell + \psi_m)}
\\
C_\ell &= \Theta_\Delta\le(\A(p_\ell)- \A(p^\star_\ell )\ri)\le( \prod_{k:\ k\neq \ell}
\frac {
 \Theta_\Delta\le(\A(p_\ell)- \A(p^\star_k )\ri) \Theta_\Delta\le(\A(p^\star_\ell)- \A(p_k )\ri)
}{ \Theta_\Delta\le(\A(p_\ell)- \A(p_k )\ri) \Theta_\Delta\le(\A(p^\star_\ell)- \A(p^\star_k )\ri)}\ri)^\frac 1 2
\ee
where the square root is defined by 
$
{\rm e}^{i\pi \mathbb B_{\ell m}}
$ 
in terms of the determination chosen in \eqref{higherB}.
If we omit a $1$ in the $(g+\ell)^{\mathrm{th}}$ position in the half-period shift, then the theorem holds but with the $\ell^{\mathrm{th}}$ row and column of the matrix $\mathbb G$ removed.
\et
The proof is an exercise following the same steps as in the proof of Theorem \ref{main}, using the general Fay identity
\be
\det &\le[\frac {\Theta\le(\A(p_\ell) - \A(q_m) + {\bf e} \ri)}{\Theta_\Delta\le(\A(p_\ell) - \A(q_m) \ri) \Theta\le({\bf e} \ri)}\ri]_{\ell, m=1}^K
=\nn\\
&=
\frac{\Theta\le(\bs A+ {\bf e} \ri)}{\Theta\le( {\bf e} \ri)}
\frac{\prod_{\ell<m} \Theta_\Delta\le(\A(p_\ell) - \A(p_m)\ri)\Theta_\Delta\le(\A(q_m) - \A(q_\ell)\ri)}{\prod_{\ell, m=1}^K \Theta_\Delta\le(\A(p_\ell) - \A(q_m)\ri)},\\
&\bs A := \sum_{\ell=1}^K \le(\A(p_\ell) - \A(q_\ell)\ri).
\ee

\section{Average and convergence in probability}
\label{sect-aver}
We now consider the following scenario where the initial phases of the degenerating curve are tuned in a random way in the part of the Jacobian associated to the degenerating cycles. In other words we consider a probability ensemble where the probability space is $\mathbb O =(S^1)^N$ representing the choices of phases associated with the gaps adjacent to the degenerating bands.
We want to prove that, in probability, the solution converges to the deterministic solution (uniformly for $(x,t)$ in compact sets) given by the elliptic cnoidal background. The Proposition below is the key estimate from which the Theorem \ref{convprob} about the convergence in probability follows easily.
From the  point of view of the geometry of the  (real section of the) Jacobian of the degenerating curve, together with Thm. \ref{main}, it shows that the probability of seeing a disturbance is an exceedingly rare event, localized in a small slice of the Jacobian around the particular half-period indicated in Thm. \ref{main}.
We parametrize the $N$--torus by $[-1,1] ^N $ (modulo even integers) for convenience. 

\bp
\label{mainAVE}
Let us denote $\bs u = (1-\phi_1,1-\phi_2,\dots, 1-\phi_N,0)^T \in \R^{N+1}$,  
${\bs X} = [\bs \psi,\beta]^T\in (i\R)^{N}\times \R$  
 and all $\phi_j\in [-1,1 ]$.  Then there exist{ constants} $K,  C$ such that 
\be
\le|\Theta\le( {\bs X} - \frac 1 2 {\bs \Omega} {\bs u} \ri) -
\theta_3\le( \beta  - \mathcal A\ri)\ri|  \leq {K} \sum_{j=1}^N {\rm e}^{-\le( \frac {|\ln \varepsilon|} 2 - C\ri) |\phi_j|}
\label{Noice2}
\ee
uniformly over compact subsets of ${\bs X}\in \C^{N+1}$. Similar estimate (with different constants) holds for any finite derivative
\be
\le|\prod_{j=1}^N\pa_{X_j}^{h_j} \pa_ \beta ^h\le(\Theta\le( {\bs X} - \frac 1 2 {\bs \Omega} {\bs u} \ri) -
\theta_3\le( \beta  - \mathcal A\ri)\ri)\ri| \leq K_{\bs h} \sum_{j=1}^N {\rm e}^{-\le( \frac {|\ln \varepsilon|} 2 - C\ri) |\phi_j|}
\ee
\ep
\noindent {\bf Proof.} We trace over the steps of the proof of Theorem \ref{main}. 
With the established notations and splitting the summation integer vector ${\bs \nu} = [{\bs n}, m]^T \in \Z^{N+1}$,  $\bs 1 = (1,1,1,\dots, 1)^T\in \R^N$ and $\bs \phi = (\phi_1,\dots, \phi_N)^T$ we have 
\begin{multline}
\Theta\le( {\bs X} - \frac 1 2 {\bs \Omega} {\bs u} \ri) 
=
\sum_{m\in \Z} \sum_{\n\in \Z^N}
\exp i\pi \le[
   m^2  \Omega_{N+1,N+1} + \n^\transpose \mathbb B \n
+2 m {\bs \mu} ^{\transpose} \n + 
2\le( 
m  \beta  
+ \n^{\transpose}\bs \psi
 - 
 \frac {1}{2} \n^\transpose \mathbb B {\bs u}
 -\frac m2 \bs \mu ^{\transpose} \bs u\ri)
\ri]\nn
 \\
\begin{multlined}[.9\textwidth]
= \sum_{\n\in \Z^N}\sum_{m\in \Z} 
\exp i\pi \Bigg[  m^2  \Omega_{N+1,N+1}  +\sum_\ell (n_\ell^2-n_\ell(1-\phi_\ell) )\mathbb B_{\ell \ell}
+ \sideset{}{'}\sum_{\ell ,k} n_\ell (n_k-1+\phi_k) \mathbb B_{\ell k} \\
 + 2m\le((\bs \mu - \bs 1/2 +\bs\phi /2)^{\transpose} \n +  \beta  \ri)+ 2 \n^{\transpose} \bs \psi  \Bigg]
\end{multlined} \\
= \sum_{\n\in \Z^N}
\exp i\pi \le[ \sum_\ell (n_\ell^2-n_\ell(1-\phi_\ell))\mathbb B_{\ell \ell}
+ \sideset{}{'}\sum_{\ell ,k} n_\ell (n_k-1+\phi_k) \mathbb B_{\ell k} 
+ 2 \n^{\transpose} \bs \psi 
\ri]\theta_3\le((\bs \mu - \frac{\bs 1}{2} - \frac{1}{2} \bs\phi^ \transpose )\n +  \beta  \ri)
\end{multline}
Now we use that  $\mathbb B_{\ell\ell} = i \ln \frac 1 \varepsilon + \mathcal O(1)$ as $\varepsilon\to 0_+$ and that all other entries have a finite limit $\mathbb B_{\ell m } \to \mathbb B_{\ell m}^0$.
Consider the quadratic form given by the off diagonal matrix $\mathbb B'=[\mathbb B_{\ell, m}]_{\ell\neq m}$: this is a family, depending on $\varepsilon$, of quadratic form and hence bounded uniformly by the norm (here below $\bs a = \bs 1 - \bs \phi$):
\be
\le|  i\pi(\bs n^\transpose \mathbb B'(\bs n + \bs a)+ 2\bs n^\transpose\bs \psi) \ri|< C_N \|\bs n\|^2 + D_N
\ee
where $C_N$ is a positive constant and  $D_N= D_N (\bs a, \bs \psi)$ is uniformly bounded (as long as $\bs \psi$ is in a compact set).
On the other hand we have that, for any $\phi\in (0,1)$, 
\be
n^2 - n(1 - \phi) = \le(n - \frac {1-\phi}2\ri)^2 - \frac {(1-\phi)^2}4 \geq 
\frac{|\phi|}2  + \frac {n^2}2-\frac 1 2 , \ \ \ \forall n\in \Z\setminus \{0\}.
\ee
We thus have the estimate
  with $\Lambda = \ln 1/\varepsilon$, 
\bea
\le| \Theta\le( {\bs X} - \frac 1 2 {\bs \Omega} {\bs u} \ri)  - \theta_3( \beta ) \ri|& \leq 
{\rm e}^{D_N}\max_{ \beta \in \R} \le|\theta_3\le(  \beta  \ri)\ri| 
\sum_{\n\in \Z^N\setminus {\bs 0}}
\prod_{\ell=1}^N \exp \le[ -\le(\frac \Lambda  2-C_N\ri) (n_\ell^2 - 1 + |\phi_\ell|)  (1-\delta_{n_\ell})   
\ri]\label{B7}
\eea
where $\delta_{n_\ell} =1$ if $n_\ell =0$ and $\delta_{n_\ell} =0$ if $n_\ell\neq 0$, and 
 we have used that $\bs \mu $  is a real vector. 
 
 Consider the expression: 
 \be
 \label{B8}
 F_{\rho} (\Lambda,\phi):= 
\sum_{|n|\geq \rho}  \exp \le[ -\le(\frac \Lambda  2-C_N\ri) (n^2 - 1 + |\phi|) (1-\delta_{n})  
\ri], \ \ \ \rho = 0,1.
\ee
We have $F_0(\Lambda,\phi) =  1+F_1(\Lambda,\phi)  $  where  
\be
\label{B9}
F_1(\Lambda, \phi) = \exp \le[ -\le(\frac \Lambda  2-C_N\ri) |\phi|\ri] \sum_{|n|\geq 1}  \exp \le[ -\le(\frac \Lambda  2-C_N\ri) (n^2 - 1) 
\ri]
\ee
and  the last series can be estimated by the integral test to provide a convergent sum as long as $\Lambda> C_N$ (recall that $\Lambda = \ln 1/\varepsilon\to + \infty$). Hence we have the uniform estimate
 \be
 F_1(\Lambda,\phi)\leq \tilde K {\rm e}^{-\le(\frac \Lambda 2 - C\ri)|\phi|}.
 \ee
Now we can rewrite the summation in \eqref{B7} as follows:
 \be
\sum_{\n\in \Z^N \setminus {\bs 0} } &
\prod_{\ell=1}^N \exp \le[ -\le(\frac \Lambda  2-C_N\ri) (n_\ell^2 - 1 + |\phi_\ell|)  (1-\delta_{n_\ell})   
\ri] \cr
&=
\sum_{\n\in \Z^N}
\prod_{\ell=1}^N \exp \le[ -\le(\frac \Lambda  2-C_N\ri) (n_\ell^2 - 1 + |\phi_\ell|)  (1-\delta_{n_\ell})   
\ri] -1 \cr
&= \prod_{j=1}^N F_0(\Lambda, \phi_j) -1= \prod_{j=1}^N \le(1 + F_1(\Lambda, \phi_j)\ri) -1
 \ee
 and thus we can estimate (for some $K>0$)
\be
\le| \Theta\le( {\bs X} - \frac 1 2 {\bs \Omega} {\bs u} \ri)  - \theta_3( \beta ) \ri|& \leq K\sum_{\ell=1}^N {\rm e}^{-\le(\frac \Lambda 2 - C\ri)|\phi_\ell|}.
\ee 
 Consider now a derivative and follow the same initial steps:
 \begin{multline}
 \nn
\prod_{j=1}^N\le(\frac{\pa_{X_j}}{2i\pi}\ri)^{k_j} \le({\pa_ \beta }\ri)^s\Theta\le( {\bs X} - \frac 1 2 {\bs \Omega} {\bs u} \ri) 
=\\
=
(2i\pi)^s\sum_{m\in \Z} \sum_{\n\in \Z^N}
\prod_{j=1}^N n_j^{k_j} m^s\exp i\pi \le[
   m^2 \Omega_{N+1,N+1} + \n^\transpose \mathbb B \n
+2 m {\bs \mu} ^{\transpose} \n + 
2\le( 
m  \beta  
+ \n^{\transpose}\bs \psi 
 - 
 \frac{1}{2} \n^\transpose \mathbb B {\bs u}
 -\frac m2 \bs \mu ^{\transpose} \bs u\ri)
\ri]\nn
=\\
= \sum_{\n\in \Z^N}\prod_{j=1}^N n_j^{k_j}
\exp i\pi\le[ \sum_\ell (n_\ell^2-n_\ell(1-\phi_\ell))\mathbb B_{\ell \ell}
+ \sum'_{\ell ,k} n_\ell (n_k-1+\phi_k) \mathbb B_{\ell k} 
+ 2 \n^{\transpose} \bs \psi 
\ri]\theta_3^{(s)}\le((\bs \mu - \frac {\bs 1}{2} - \frac{1}{2}\bs\phi)^\transpose \n +  \beta  \ri)
\end{multline}
If all the $k_j$'s are zero (and $s\neq 0$), then the estimate proceeds exactly as before.
If at least one of the $k_j$'s is nonzero, since the quadratic form at the exponent has its maximum at $\bs n=\bs 0$, we have the equivalent of \eqref{B7}
\begin{multline}
\le|
 \prod_{j=1}^N\le(\frac{\pa_{X_j}}{2i\pi}\ri)^{k_j} \le({\pa_ \beta }\ri)^s\Theta\le( {\bs X} - \frac 1 2 {\bs \Omega} {\bs u} \ri)   \ri| \leq \\
e^{ D_N} \max_{ \beta \in \R} \le|\theta_3^{(s)}\le(  \beta  \ri)\ri| \sideset{}{'}\sum_{\n\in \Z^N}
\prod_{\ell=1}^Nn_\ell^{k_\ell} \exp \le[ -\le(\frac \Lambda  2-C_N\ri) (n_\ell^2 - 1 + |\phi_\ell|)  (1-\delta_\ell)  
\ri]
\end{multline}
Then the same reasoning around \eqref{B8}, \eqref{B9} applies (with different constants).
 \QED
 \bt
 \label{convprob}
 Let $\bs \Phi$ denote the $N$--dimensional real torus of the phases $\bs\phi \in [-1,1)^N$, with the normalized (unit) volume $\frac 1{2^N}\d^N \bs \phi$,  considered  as a probability space. 
 Let $u_{N+1}(x,t; \vec \phi; \varepsilon)$ denote the $N+1$-gap solution of the KdV equation with the initial phases $\phi_1,\dots, \phi_N$, thought of as a random variable (depending on $x,t$). Denote by $u_1(x,t) = 2\pa_x^2 \ln \theta_3\le(\frac {x}{4i{\varpi_3}}\ri)$ the deterministic cnoidal stationary wave solution.  Then 
 \be
 u_{N+1}(x,t; \vec \phi; \varepsilon) \to u_1(x,t) 
 \ee
 in probability, uniformly for $(x,t)$ in compact sets.
 \et
 \noindent 
 {\bf Proof.}
 Let $(x,t)$ belong to a compact set and consider  the estimate of Prop. \ref{mainAVE}.  Then  for some constant $K>0$ (which depends only on the chosen compact set),   
 \be
 | u_{N+1}(x,t; \vec \phi; \varepsilon) -u_1(x,t) |\leq K\sum_{j=1}^{N} {\rm e}^{-
 \le( \frac {|\ln \varepsilon|} 2 - C\ri)|\phi_j|},
 \ee
 for all $(x,t)$ in that set. 
 The integral of the latter function on the torus is $\mathcal O\le(\frac 1 {|\ln \varepsilon|}\ri)$ which tends to zero as $\varepsilon\to 0_+$. \QED

 \end{document}